\documentclass[a4paper,fleqn,usenatbib]{mnras}

\usepackage{newtxtext,newtxmath}

\usepackage[T1]{fontenc}
\usepackage{ae,aecompl}

\usepackage{graphicx}	
\usepackage{amsmath}	
\usepackage{amssymb}	

\usepackage{color}
\usepackage{natbib,twoopt}
\usepackage{hyperref} 
\usepackage{booktabs}
\usepackage{multirow}
\definecolor{mygray}{gray}{0.7}

\newcommand{\Teff}{\ensuremath{T_{\mathrm{eff}}}}
\newcommand{\logg}{\ensuremath{\log\,g}}
\newcommand{\pun}[1]{\mbox{\rm\,#1}} 

\newcommand{\HARPS}{{\sf HARPS}}

\newcommand{\vconv}{\ensuremath{v_\mathrm{conv}}}

\usepackage{breakurl}
\usepackage{longtable}
\usepackage{ragged2e}

\title[Spectroscopic and astrometric radial velocities]{Spectroscopic and astrometric radial velocities: Hyades as a benchmark\thanks{Based on observations made with ESO telescopes at the La Silla Observatory under programme ID 094.D-0596(A).}}

\author[I. C. Le\~ao et al.]{
I.~C.~Le\~ao$^{1}$\thanks{E-mail: izan@dfte.ufrn.br},
L.~Pasquini$^{2}$,
Ludwig, H.-G.$^{3}$,
and J.~R.~de~Medeiros$^{1}$
\\
$^{1}$Universidade Federal do Rio Grande do Norte, Natal, Brazil\\
$^{2}$ESO, Karl-Schwarzschild-Stra{\ss}e 2, 85748 Garching bei M\"unchen, Germany\\
$^{3}$Zentrum f\"ur Astronomie der Universit\"at Heidelberg, Landessternwarte, K\"onigstuhl 12, D-69117 Heidelberg, Germany\\
}

\date{Accepted XXX. Received YYY; in original form ZZZ}

\pubyear{2017}

\begin{document}
\label{firstpage}
\pagerange{\pageref{firstpage}--\pageref{lastpage}}
\maketitle

\begin{abstract}
We study the accuracy of spectroscopic RVs by comparing 
spectroscopic and astrometric RVs for stars of the Hyades open cluster.
Rather accurate astrometric RVs are available for the Hyades' stars, based on Hipparcos and on the first
{\em Gaia} data release. We obtained HARPS spectra for a large sample of Hyades stars, and we homogeneously analysed them.
After cleaning the sample from binaries, RV variables, and outliers, 71 stars remained.
The distribution of the observed RV difference (spectroscopic -- astrometric) is
skewed
and depends on the star right ascension. This is consistent with the Hyades cluster rotating
at $42.3$~m~s$^{-1}$pc$^{-1}$.
The two Hyades giants in the sample show, as predicted by gravitational redshift (GR), 
a spectroscopic RV that is blue-shifted with respect to the dwarfs, and the empirical GR slope is of $626 \pm 131$~m~$s^{-1}$,
in agreement with the theoretical prediction.
The difference between spectroscopic and astrometric RVs is very close to zero, within the uncertainties.
In particular, the mean difference is of $-33$~m~s$^{-1}$ and the median is of $-16$~m~s$^{-1}$
when considering the {\em Gaia}-based RVs (corrected for cluster rotation), with a $\sigma$ of $347$~m~s$^{-1}$, very close to the
expected cluster velocity dispersion.
We also determine a new value of the cluster centroid spectroscopic RV: $39.36 \pm0.26$~km~s$^{-1}$. 
The spectroscopic RV measurements are expected, from simulations, to depend on stellar rotation, 
but our data do not confirm these predictions.
We finally discuss the other phenomena that can influence the RV difference, such as cluster expansion,
stellar activity, general relativity, and Galactic potential. Clusters within the reach of current telescopes
are expected to show differences of several hundreds m~s$^{-1}$, depending on their position in the Galaxy.
\end{abstract}




\section{Introduction} \label{sec:intro}

The measurement of the Doppler shift between the absorption lines of a star with respect to a reference spectrum is commonly 
dubbed the stellar ``radial velocity'' (RV), but in reality these measurements are affected by a number of effects, 
that are not easy to measure and new definitions of RV have been recently adopted by IAU, one for spectroscopic mesurements, one for 
astrometric ones \citep[][for a full discussion]{Lindegren03}.

\citet{Dravins99} discuss three methods to obtain astrometric RVs, 
one of which is suitable for moving groups and clusters. It is based on the concept 
that the angular extent of a cluster changes with time because, thanks to its RV, its 
distance will change. Unfortunately, precise astrometric RVs are only available 
for a small number of stars and for a few open clusters {\citep[][M02 hereafter]{Madsen02}}.

The early comparison between spectroscopic and astrometric RVs for the stars of the Hyades has 
shown a fair agreement between the two measurements, and a potential dependence of the RV difference
on the stellar spectral type and rotational velocity (M02). 
Amongst the effects affecting the spectroscopic measurements but not the 
astrometric ones, GR is likely the most prominent, since it affects differentially non-degenerate stars of 
the same cluster of up to several hundreds m~s$^{-1}$ and white dwarfs up to km~s$^{-1}$. On a previous study of giants and dwarfs in 
the open cluster M67, \citet{Pasquini11} did not find evidence for GR, when measuring the RVs 
by using cross correlation techniques and the same digital mask for all the stars. The authors generated synthetic spectral 
lines with 3D atmospheres models, and found that the increased photospheric blue shift in solar-type main sequence stars 
roughly compensates the lower GR effect in evolved stars. 
These first results show a rather intriguing situation, and makes the problem worth to be studied in more detail.

We present a comparison between spectroscopic RV data collected with the HARPS spectrograph \citep{Mayor03} and astrometric data from M02.
While developing this work, the {\em Gaia} Mission provided its first data release \citep[][G17 henceforth]{gaia17}
with new astrometric data and a potentially more accurate estimate of the space velocity of the cluster, which is 
the only information we need to estimate astrometric RVs.
Some {\em Gaia} data are not available for Hyades stars (mostly bright) because of reduction issues that may be fixed in the future,
but they can be replaced by other data (specifically the equatorial coordinates $\alpha$ and $\delta$) for the calculation of astrometric RVs (see Sect.~\ref{sec:astromrvs} for more details).

The main questions we address in this work is to which extent astrometric and spectroscopic RV measurements agree, 
and to which extent the difference between spectroscopic and astrometric RVs are affected by stellar activity, 
GR, atmospheric 3D effects, cluster expansion, galactic potential.
We note that, since the results of two independent methods (Doppler vs. astrometric) are compared, 
 the accuracy of the measurements matters, rather than their precision. 
 This is a rare case, since, in general, the RV variations with time (and therefore precision) 
 is what most users measure. 

Relevant data about the Hyades are presented in Sect.~\ref{sec:hyades}, together with our HARPS observations and
the estimate of the physical stellar parameters used in our analysis. A first analysis of the HARPS RV measurements 
is presented in Sec.~\ref{sec:analysis}, where they are compared with astrometric and spectroscopic RVs from the literature.
A more detailed discussion between HARPS and astrometric RVs is provided in Sect.~\ref{sec:rvastrovsspec}, 
where major and minor effects that may distort these measurements are discussed and the results 
of our 3D simulations are applied.
 The major effects are analysed in more detail in Sect.~\ref{sec:maineffects}, where an accurate GR measurement is obtained for the Hyades stars. A HARPS-based cluster RV is determined in Sect.~\ref{sec:averagervs} by considering the spatial distribution of the Hyades members.
Finally, our conclusions are summarized in Sect.~\ref{sec:conclusion}.

\section{Hyades: the benchmark test} \label{sec:hyades}

The Hyades is a relatively young cluster, with an age of 
$\sim$~625~$\pm$~50~$\rm{Myr}$ (P98), and it is the closest open cluster 
to the Sun \citep[$\sim$~46.5~pc,][]{vanLeeuwen09}. 
Several astrometric studies of the Hyades exist, based on Hipparcos measurements (P98, M02, \citealt{deBruijne01}, \citealt{vanLeeuwen07}), and recently G17 provided a
first {\em Gaia} solution. 
For this work we use as a baseline the M02 study, that provided a clean sample, 
astrometric RVs for al stars and compared them with the spectroscopic ones. 

The Hyades metallicity has been studied in detail ($[Fe/H] = 0.18$~dex, \citealt{DutraFerreira16})
and the good parallaxes measurements coupled to the other observations make it possible to compute rather precise stellar parameters.
Hyades stars have also been targets of planet searches using RVs \citep{Cochran02,Paulson04} and one of the four Hyades giants was reported to host a giant planet. 
More recently, \cite{Quinn14} reported the discovery of the first hot Jupiter 
orbiting a K dwarf HD~285507 in this cluster. The relevance of these studies for the present work is mainly 
that \citet{Paulson04} constrained the spectroscopic RV jitter of typical solar-type stars 
in the Hyades to less than 40~m~s$^{-1}$.
This jitter provides therefore a fundamental limit to the accuracy that can be obtained from the measurement of 
a single spectrum. Even if the HARPS RV precision is of a few m~s$^{-1}$ or less, 
the accuracy of the spectroscopic RV will be limited by this jitter, having only one observation/star. 

A total of 218 stars, selected from the list of P98, were observed,
out of which 131 match with the list of M02, 
which is cleaned for membership and some binarity. 

\begin{table*}
 \caption{Selected objects in the Hyades cluster.} 
 \label{table:main}
 {\centering
 \setlength{\tabcolsep}{0.54em}
 \begin{tabular}{cccccccc|cccc}
\toprule
\# & HIP & SpType & $V_{\rm mag}$ & $(B-V)$ & $RV_{\rm P98}^{\rm Spectr.}$ & $RV_{\rm M02}^{\rm Astrom.}$ & $RV_{\rm G17}^{\rm Astrom.}$ & $RV_{\rm HARPS}$ & FWHM & $S/N$ & Flag \\
 & & & (mag) & (mag) & (km~s$^{-1}$) & (km~s$^{-1}$) & (km~s$^{-1}$) & (km~s$^{-1}$) & (km~s$^{-1}$) & & \\
\midrule
1 & 13600 & K0 & 8.83 & 0.704 $\pm$ 0.023 & 30.41 $\pm$ 0.23 & 27.73 $\pm$ 0.58 & 27.59 $\pm$ 0.13 & 31.1452 $\pm$ 0.0004 & 7.0 & 116 & 0	\\
2 & 16529 & G5 & 8.88 & 0.844 $\pm$ 0.000 & 32.72 $\pm$ 0.17 & 32.47 $\pm$ 0.59 & 32.34 $\pm$ 0.14 & 32.0867 $\pm$ 0.0007 & 7.5 & 116 & 0	\\
3 & 16908 & G5 & 9.35 & 0.917 $\pm$ 0.001 & 33.56 $\pm$ 0.21 & 33.39 $\pm$ 0.59 & 33.26 $\pm$ 0.14 & 31.6168 $\pm$ 0.0007 & 7.4 & 116 & 0	\\
4 & 18327 & K2 & 8.99 & 0.895 $\pm$ 0.001 & 36.79 $\pm$ 0.13 & 36.03 $\pm$ 0.59 & 35.90 $\pm$ 0.15 & 36.4212 $\pm$ 0.0010 & 8.5 & 97 & 0	\\
5 & 19098 & K2 & 9.31 & 0.890 $\pm$ 0.008 & 37.61 $\pm$ 0.05 & 37.16 $\pm$ 0.59 & 37.03 $\pm$ 0.16 & 36.8190 $\pm$ 0.0007 & 8.3 & 98 & 0	\\
6 & 19148 & G0 & 7.85 & 0.593 $\pm$ 0.005 & 38.04 $\pm$ 0.17 & 37.44 $\pm$ 0.59 & 37.31 $\pm$ 0.16 & 37.5821 $\pm$ 0.0009 & 10.7 & 131 & 0	\\
7 & 19365 & G0IV & 8.27 & 0.637 $\pm$ 0.002 & 37.92 $\pm$ 0.15 & 35.57 $\pm$ 0.59 & 35.45 $\pm$ 0.15 & 38.8555 $\pm$ 0.0012 & 10.0 & 122 & 0	\\
8 & 19781 & G5V & 8.45 & 0.693 $\pm$ 0.011 & 39.24 $\pm$ 0.06 & 38.43 $\pm$ 0.60 & 38.30 $\pm$ 0.16 & 38.1808 $\pm$ 0.0013 & 9.3 & 97 & 0	\\
9 & 19796 & F8V & 7.11 & 0.514 $\pm$ 0.008 & 38.50 $\pm$ 0.15 & 38.65 $\pm$ 0.60 & 38.52 $\pm$ 0.16 & 38.2816 $\pm$ 0.0042 & 22.6 & 130 & 0	\\
10 & 20146 & G8V & 8.47 & 0.721 $\pm$ 0.016 & 38.80 $\pm$ 0.08 & 38.66 $\pm$ 0.60 & 38.53 $\pm$ 0.16 & 38.0862 $\pm$ 0.0010 & 9.3 & 124 & 0	\\
11 & 20205 & G8III & 3.65 & 0.981 $\pm$ 0.012 & 39.28 $\pm$ 0.11 & 38.91 $\pm$ 0.60 & 38.79 $\pm$ 0.16 & 38.5485 $\pm$ 0.0007 & 8.5 & 116 & 0	\\
12 & 20480 & G9V & 8.84 & 0.758 $\pm$ 0.003 & 39.24 $\pm$ 0.24 & 38.56 $\pm$ 0.60 & 38.44 $\pm$ 0.16 & 38.7314 $\pm$ 0.0009 & 8.2 & 105 & 0	\\
13 & 20492 & K1V & 9.11 & 0.855 $\pm$ 0.018 & 40.29 $\pm$ 0.06 & 39.38 $\pm$ 0.60 & 39.26 $\pm$ 0.16 & 39.5077 $\pm$ 0.0010 & 8.4 & 100 & 0	\\
14 & 20557 & F8V & 7.13 & 0.518 $\pm$ 0.014 & 38.94 $\pm$ 0.13 & 38.59 $\pm$ 0.60 & 38.47 $\pm$ 0.16 & 38.2739 $\pm$ 0.0038 & 17.2 & 102 & 0	\\
15 & 20614 & F4V & 5.97 & 0.378 $\pm$ 0.014 & 36.60 $\pm$ 1.20 & 39.06 $\pm$ 0.60 & 38.94 $\pm$ 0.16 & 31.9955 $\pm$ 0.0074 & 9.4 & 175 & 0	\\
16 & 20826 & F8 & 7.49 & 0.560 $\pm$ 0.007 & 40.22 $\pm$ 0.21 & 39.99 $\pm$ 0.60 & 39.86 $\pm$ 0.17 & 39.6935 $\pm$ 0.0020 & 13.6 & 127 & 0	\\
17 & 20850 & K0V & 9.02 & 0.839 $\pm$ 0.007 & 40.94 $\pm$ 0.08 & 39.89 $\pm$ 0.60 & 39.77 $\pm$ 0.16 & 40.2936 $\pm$ 0.0009 & 8.0 & 98 & 0	\\
18 & 20889 & K0III & 3.53 & 1.014 $\pm$ 0.006 & 39.37 $\pm$ 0.06 & 39.39 $\pm$ 0.60 & 39.27 $\pm$ 0.16 & 38.4354 $\pm$ 0.0004 & 9.0 & 199 & 0	\\
19 & 20949 & G5 & 9.19 & 0.766 $\pm$ 0.002 & 39.02 $\pm$ 0.17 & 38.12 $\pm$ 0.59 & 38.01 $\pm$ 0.15 & 38.6972 $\pm$ 0.0011 & 7.5 & 76 & 0	\\
20 & 20978 & K1V & 9.08 & 0.865 $\pm$ 0.013 & 40.97 $\pm$ 0.06 & 39.83 $\pm$ 0.60 & 39.71 $\pm$ 0.16 & 40.2908 $\pm$ 0.0017 & 8.6 & 59 & 0	\\
21 & 21099 & G8V & 8.59 & 0.734 $\pm$ 0.014 & 40.62 $\pm$ 0.08 & 39.51 $\pm$ 0.60 & 39.39 $\pm$ 0.16 & 39.8309 $\pm$ 0.0009 & 8.4 & 109 & 0	\\
22 & 21741 & K0V & 9.40 & 0.811 $\pm$ 0.000 & 41.34 $\pm$ 0.16 & 39.75 $\pm$ 0.60 & 39.64 $\pm$ 0.16 & 40.2362 $\pm$ 0.0013 & 8.9 & 84 & 0	\\
23 & 22380 & F5 & 8.98 & 0.833 $\pm$ 0.003 & 41.62 $\pm$ 0.15 & 41.27 $\pm$ 0.60 & 41.16 $\pm$ 0.17 & 41.7660 $\pm$ 0.0010 & 8.4 & 95 & 0	\\
24 & 22422 & F8 & 7.72 & 0.578 $\pm$ 0.009 & 42.04 $\pm$ 0.14 & 41.65 $\pm$ 0.60 & 41.53 $\pm$ 0.17 & 41.5230 $\pm$ 0.0015 & 9.6 & 96 & 0	\\
25 & 22566 & F8 & 7.90 & 0.527 $\pm$ 0.008 & 42.92 $\pm$ 0.19 & 41.85 $\pm$ 0.60 & 41.73 $\pm$ 0.17 & 42.4678 $\pm$ 0.0025 & 15.2 & 126 & 0	\\
26 & 23069 & G5 & 8.89 & 0.737 $\pm$ 0.001 & 43.68 $\pm$ 0.16 & 42.48 $\pm$ 0.60 & 42.37 $\pm$ 0.17 & 42.9643 $\pm$ 0.0009 & 7.6 & 100 & 0	\\
27 & 23498 & K0 & 9.00 & 0.765 $\pm$ 0.002 & 43.51 $\pm$ 0.19 & 42.90 $\pm$ 0.60 & 42.79 $\pm$ 0.17 & 43.2023 $\pm$ 0.0012 & 8.9 & 93 & 0	\\
28 & 23750 & K0 & 8.82 & 0.730 $\pm$ 0.015 & 42.31 $\pm$ 0.18 & 42.67 $\pm$ 0.60 & 42.56 $\pm$ 0.17 & 42.8852 $\pm$ 0.0012 & 9.4 & 101 & 0	\\
29 & 24923 & K0 & 9.03 & 0.765 $\pm$ 0.029 & 43.70 $\pm$ 0.23 & 44.17 $\pm$ 0.60 & 44.07 $\pm$ 0.17 & 44.5765 $\pm$ 0.0011 & 9.1 & 107 & 0	\\
30 & 10672 & G0 & 8.55 & 0.567 $\pm$ 0.021 & 26.40 $\pm$ 0.32 & 21.30 $\pm$ 0.56 & 21.15 $\pm$ 0.12 & 27.0446 $\pm$ 0.0009 & 11.0 & 108 & 1	\\
31 & 15300 & M0 & 11.11 & 1.408 $\pm$ 0.007 & 29.84 $\pm$ 0.29 & 30.06 $\pm$ 0.58 & 29.93 $\pm$ 0.13 & 29.3413 $\pm$ 0.0031 & 7.5 & 45 & 1	\\
32 & 15563 & K7V: & 9.65 & 1.130 $\pm$ 0.015 & 30.45 $\pm$ 0.26 & 31.72 $\pm$ 0.58 & 31.58 $\pm$ 0.15 & 31.0463 $\pm$ 0.0006 & 8.2 & 99 & 1	\\
33 & 15720 & -- & 11.03 & 1.431 $\pm$ 0.004 & 28.90 $\pm$ 0.45 & 30.60 $\pm$ 0.58 & 30.47 $\pm$ 0.13 & 29.9547 $\pm$ 0.0023 & 7.2 & 59 & 1	\\
34 & 16548 & M0V: & 11.88 & 1.378 $\pm$ 0.015 & 26.60 $\pm$ 0.34 & 33.41 $\pm$ 0.58 & 33.27 $\pm$ 0.15 & 28.6738 $\pm$ 0.0029 & 7.3 & 45 & 1	\\
35 & 17766 & M1 & 10.85 & 1.340 $\pm$ 0.015 & 35.40 $\pm$ 0.25 & 35.51 $\pm$ 0.59 & 35.37 $\pm$ 0.16 & 34.9527 $\pm$ 0.0022 & 7.4 & 53 & 1	\\
36 & 18018 & K7 & 10.17 & 1.160 $\pm$ 0.040 & 35.30 $\pm$ 0.12 & 34.67 $\pm$ 0.59 & 34.55 $\pm$ 0.15 & 34.5523 $\pm$ 0.0010 & 7.6 & 85 & 1	\\
37 & 18322 & M0 & 10.12 & 1.070 $\pm$ 0.015 & 37.18 $\pm$ 0.22 & 36.32 $\pm$ 0.59 & 36.18 $\pm$ 0.16 & 36.2453 $\pm$ 0.0008 & 8.1 & 89 & 1	\\
38 & 18946 & K5 & 10.12 & 1.095 $\pm$ 0.119 & 36.93 $\pm$ 0.26 & 36.76 $\pm$ 0.59 & 36.63 $\pm$ 0.15 & 36.0967 $\pm$ 0.0008 & 7.8 & 107 & 1	\\
39 & 19082 & -- & 11.41 & 1.347 $\pm$ 0.005 & 38.33 $\pm$ 0.22 & 36.96 $\pm$ 0.59 & 36.83 $\pm$ 0.15 & 37.4701 $\pm$ 0.0028 & 7.6 & 46 & 1	\\
40 & 19207 & K5 & 10.49 & 1.180 $\pm$ 0.015 & 38.95 $\pm$ 0.23 & 37.56 $\pm$ 0.59 & 37.43 $\pm$ 0.16 & 38.1161 $\pm$ 0.0016 & 8.1 & 64 & 1	\\
41 & 19263 & K0 & 9.94 & 1.005 $\pm$ 0.012 & 38.72 $\pm$ 0.05 & 37.53 $\pm$ 0.59 & 37.41 $\pm$ 0.16 & 38.0033 $\pm$ 0.0008 & 7.8 & 105 & 1	\\
42 & 19316 & -- & 11.28 & 1.327 $\pm$ 0.004 & 38.43 $\pm$ 0.28 & 37.91 $\pm$ 0.59 & 37.78 $\pm$ 0.16 & 37.9994 $\pm$ 0.0029 & 7.7 & 44 & 1	\\
43 & 19441 & K5V & 10.10 & 1.192 $\pm$ 0.011 & 39.24 $\pm$ 0.16 & 38.15 $\pm$ 0.59 & 38.01 $\pm$ 0.16 & 38.2760 $\pm$ 0.0009 & 7.7 & 98 & 1	\\
44 & 19808 & K5 & 10.69 & 1.204 $\pm$ 0.008 & 40.51 $\pm$ 0.15 & 38.58 $\pm$ 0.60 & 38.45 $\pm$ 0.16 & 39.0411 $\pm$ 0.0020 & 7.7 & 52 & 1	\\
45 & 19834 & K2 & 11.56 & 1.363 $\pm$ 0.010 & 38.79 $\pm$ 0.36 & 38.53 $\pm$ 0.60 & 38.40 $\pm$ 0.16 & 38.2546 $\pm$ 0.0026 & 7.7 & 51 & 1	\\
46 & 19862 & M2 & 10.96 & 0.924 $\pm$ 0.301 & 38.96 $\pm$ 0.17 & 38.46 $\pm$ 0.60 & 38.34 $\pm$ 0.16 & 38.6097 $\pm$ 0.0017 & 7.2 & 58 & 1	\\
47 & 20357 & F5V & 6.60 & 0.412 $\pm$ 0.014 & 39.20 $\pm$ 0.21 & 39.20 $\pm$ 0.60 & 39.07 $\pm$ 0.16 & 39.1183 $\pm$ 0.0075 & 28.5 & 119 & 1	\\
48 & 20527 & K5.5Ve & 10.89 & 1.288 $\pm$ 0.002 & 40.64 $\pm$ 0.26 & 39.46 $\pm$ 0.60 & 39.34 $\pm$ 0.16 & 39.7234 $\pm$ 0.0019 & 7.4 & 55 & 1	\\
49 & 20605 & M0.5Ve & 11.66 & 1.408 $\pm$ 0.013 & 40.20 $\pm$ 0.36 & 39.40 $\pm$ 0.60 & 39.28 $\pm$ 0.16 & 39.9854 $\pm$ 0.0060 & 9.4 & 38 & 1	\\
50 & 20745 & M0V & 10.50 & 1.358 $\pm$ 0.002 & 41.38 $\pm$ 0.18 & 39.84 $\pm$ 0.60 & 39.71 $\pm$ 0.16 & 40.2956 $\pm$ 0.0022 & 7.6 & 56 & 1	\\
51 & 20762 & K7 & 10.48 & 1.146 $\pm$ 0.001 & 41.22 $\pm$ 0.21 & 39.83 $\pm$ 0.60 & 39.70 $\pm$ 0.16 & 40.2113 $\pm$ 0.0018 & 8.2 & 57 & 1	\\
52 & 20827 & K0 & 9.48 & 0.929 $\pm$ 0.005 & 40.46 $\pm$ 0.07 & 39.82 $\pm$ 0.60 & 39.70 $\pm$ 0.16 & 39.7185 $\pm$ 0.0008 & 8.0 & 103 & 1	\\
53 & 21138 & K5Ve & 11.02 & 1.280 $\pm$ 0.014 & 41.28 $\pm$ 0.21 & 40.13 $\pm$ 0.60 & 40.01 $\pm$ 0.16 & 40.3908 $\pm$ 0.0025 & 7.7 & 46 & 1	\\
54 & 21256 & K8 & 10.69 & 1.237 $\pm$ 0.005 & 41.39 $\pm$ 0.20 & 39.56 $\pm$ 0.60 & 39.45 $\pm$ 0.16 & 40.0236 $\pm$ 0.0013 & 7.6 & 74 & 1	\\
55 & 21261 & -- & 10.74 & 1.197 $\pm$ 0.004 & 41.43 $\pm$ 0.15 & 39.89 $\pm$ 0.60 & 39.77 $\pm$ 0.16 & 40.1629 $\pm$ 0.0015 & 8.1 & 69 & 1	\\
\bottomrule
 \end{tabular} \\}
 \small
 \begin{justify}
\textbf{Notes.}
Columns are HIP: Hipparcos number; SpType: spectral type from the Hipparcos catalogue; $V_{\rm mag}$: visual magnitude; $(B-V)$: colour index; $RV_{\rm P98}^{\rm Spectr.}$: spectroscopic RV from P98; $RV_{\rm M02}^{\rm Astrom.}$: astrometric RV from M02; $RV_{\rm G17}^{\rm Astrom.}$: astrometric RV computed in this work from G17 data; $RV_{\rm HARPS}$: our HARPS RV measurements without zero point correction, from which we subtracted the HARPS mask zero point of $102$~m~s$^{-1}$ for our analysis (see text); FWHM: full width at half maximum of the HARPS CCF; $S/N$: singal-to-noise ratio at 490~$\mu$m; Flag: quality flag. Flags are 0 and 1 for acceptable HARPS data with higher- and lower-quality CCF shape, respectively; ``BY'' for BY~Dra variables with acceptable CCF shape; ``SB'' for spectroscopic binaries; and ``x'' for excluded targets (because of bad-quality CCF).
 \end{justify}
\end{table*}

\setcounter{table}{0}
\begin{table*}
 \caption{Continued.} 
 \centering
 \setlength{\tabcolsep}{0.54em}
 \begin{tabular}{cccccccc|cccc}
\toprule
\# & HIP & SpType & $V_{\rm mag}$ & $(B-V)$ & $RV_{\rm P98}^{\rm Spectr.}$ & $RV_{\rm M02}^{\rm Astrom.}$ & $RV_{\rm G17}^{\rm Astrom.}$ & $RV_{\rm HARPS}$ & FWHM & $S/N$ & Flag \\
 & & & (mag) & (mag) & (km~s$^{-1}$) & (km~s$^{-1}$) & (km~s$^{-1}$) & (km~s$^{-1}$) & (km~s$^{-1}$) & & \\
\midrule
56 & 21723 & K5 & 10.04 & 1.073 $\pm$ 0.005 & 42.50 $\pm$ 0.19 & 41.08 $\pm$ 0.60 & 40.96 $\pm$ 0.17 & 41.6793 $\pm$ 0.0010 & 7.9 & 89 & 1	\\
57 & 22177 & -- & 10.92 & 1.277 $\pm$ 0.005 & 43.16 $\pm$ 0.25 & 41.71 $\pm$ 0.60 & 41.59 $\pm$ 0.17 & 41.9002 $\pm$ 0.0021 & 7.5 & 50 & 1	\\
58 & 22253 & K2III & 10.69 & 1.112 $\pm$ 0.004 & 41.78 $\pm$ 0.23 & 40.40 $\pm$ 0.60 & 40.29 $\pm$ 0.16 & 40.6158 $\pm$ 0.0014 & 8.3 & 71 & 1	\\
59 & 22271 & K7III & 10.61 & 1.174 $\pm$ 0.005 & 40.30 $\pm$ 0.17 & 39.76 $\pm$ 0.60 & 39.65 $\pm$ 0.16 & 39.5603 $\pm$ 0.0013 & 7.2 & 70 & 1	\\
60 & 22654 & K0 & 10.29 & 1.070 $\pm$ 0.015 & 42.88 $\pm$ 0.25 & 41.48 $\pm$ 0.60 & 41.37 $\pm$ 0.17 & 41.6956 $\pm$ 0.0012 & 8.2 & 76 & 1	\\
61 & 23312 & K2 & 9.71 & 0.957 $\pm$ 0.047 & 42.21 $\pm$ 0.40 & 42.94 $\pm$ 0.60 & 42.82 $\pm$ 0.17 & 43.0503 $\pm$ 0.0007 & 7.6 & 110 & 1	\\
62 & 13806 & G5 & 8.92 & 0.855 $\pm$ 0.008 & 26.62 $\pm$ 0.21 & 26.86 $\pm$ 0.57 & 26.74 $\pm$ 0.12 & 26.3762 $\pm$ 0.0010 & 8.6 & 99 & BY	\\
63 & 13976 & G5 & 7.97 & 0.926 $\pm$ 0.015 & 28.35 $\pm$ 0.18 & 28.60 $\pm$ 0.57 & 28.46 $\pm$ 0.14 & 28.9233 $\pm$ 0.0003 & 7.8 & 119 & BY	\\
64 & 19786 & G0 & 8.05 & 0.640 $\pm$ 0.001 & 39.32 $\pm$ 0.14 & 38.57 $\pm$ 0.60 & 38.44 $\pm$ 0.16 & 38.4077 $\pm$ 0.0011 & 9.7 & 122 & BY	\\
65 & 19793 & G3V & 8.05 & 0.657 $\pm$ 0.007 & 38.21 $\pm$ 0.23 & 37.31 $\pm$ 0.59 & 37.19 $\pm$ 0.15 & 37.6314 $\pm$ 0.0011 & 10.0 & 138 & BY	\\
66 & 19934 & K0V & 9.14 & 0.813 $\pm$ 0.002 & 38.46 $\pm$ 0.19 & 37.80 $\pm$ 0.59 & 37.68 $\pm$ 0.16 & 38.2191 $\pm$ 0.0006 & 7.5 & 136 & BY	\\
67 & 20082 & K3V & 9.57 & 0.980 $\pm$ 0.005 & 39.64 $\pm$ 0.08 & 38.72 $\pm$ 0.60 & 38.59 $\pm$ 0.16 & 38.9685 $\pm$ 0.0008 & 7.4 & 98 & BY	\\
68 & 20130 & G9V & 8.62 & 0.745 $\pm$ 0.005 & 39.58 $\pm$ 0.06 & 38.34 $\pm$ 0.60 & 38.22 $\pm$ 0.16 & 38.7177 $\pm$ 0.0006 & 8.2 & 160 & BY	\\
69 & 20237 & G0V & 7.46 & 0.560 $\pm$ 0.014 & 38.81 $\pm$ 0.18 & 38.56 $\pm$ 0.60 & 38.44 $\pm$ 0.16 & 38.3962 $\pm$ 0.0018 & 15.4 & 168 & BY	\\
70 & 20485 & K5V & 10.47 & 1.231 $\pm$ 0.009 & 39.30 $\pm$ 0.21 & 39.27 $\pm$ 0.60 & 39.15 $\pm$ 0.16 & 38.4992 $\pm$ 0.0019 & 8.1 & 59 & BY	\\
71 & 20563 & K4V & 9.99 & 1.050 $\pm$ 0.011 & 39.95 $\pm$ 0.16 & 39.12 $\pm$ 0.60 & 39.00 $\pm$ 0.16 & 39.4498 $\pm$ 0.0008 & 7.8 & 111 & BY	\\
72 & 20577 & G2V & 7.79 & 0.599 $\pm$ 0.015 & 38.80 $\pm$ 0.08 & 39.27 $\pm$ 0.60 & 39.15 $\pm$ 0.16 & 37.7569 $\pm$ 0.0013 & 10.8 & 136 & BY	\\
73 & 20741 & G8V & 8.10 & 0.664 $\pm$ 0.008 & 40.23 $\pm$ 0.28 & 39.50 $\pm$ 0.60 & 39.38 $\pm$ 0.16 & 38.3392 $\pm$ 0.0011 & 8.0 & 89 & BY	\\
74 & 20815 & F8V & 7.41 & 0.537 $\pm$ 0.015 & 39.32 $\pm$ 0.24 & 39.71 $\pm$ 0.60 & 39.58 $\pm$ 0.16 & 39.3271 $\pm$ 0.0026 & 14.3 & 108 & BY	\\
75 & 20899 & G2V & 7.83 & 0.609 $\pm$ 0.010 & 39.99 $\pm$ 0.16 & 39.65 $\pm$ 0.60 & 39.53 $\pm$ 0.16 & 39.2047 $\pm$ 0.0013 & 11.2 & 137 & BY	\\
76 & 20951 & K0V & 8.95 & 0.831 $\pm$ 0.003 & 40.70 $\pm$ 0.06 & 39.65 $\pm$ 0.60 & 39.53 $\pm$ 0.16 & 39.9421 $\pm$ 0.0010 & 7.6 & 82 & BY	\\
77 & 21317 & G1V & 7.90 & 0.631 $\pm$ 0.014 & 40.78 $\pm$ 0.16 & 40.38 $\pm$ 0.60 & 40.26 $\pm$ 0.16 & 40.4391 $\pm$ 0.0012 & 9.9 & 120 & BY	\\
78 & 19554 & F4V & 5.71 & 0.360 $\pm$ 0.012 & 36.60 $\pm$ 1.20 & 38.28 $\pm$ 0.59 & 38.14 $\pm$ 0.16 & 36.9090 $\pm$ 0.0045 & 21.8 & 152 & SB	\\
79 & 19870 & G4V & 7.83 & 0.705 $\pm$ 0.003 & 38.46 $\pm$ 0.12 & 37.88 $\pm$ 0.59 & 37.75 $\pm$ 0.16 & 20.1114 $\pm$ 0.0015 & 8.2 & 123 & SB	\\
80 & 20019 & G8V & 8.32 & 0.756 $\pm$ 0.004 & 38.18 $\pm$ 0.13 & 38.56 $\pm$ 0.60 & 38.44 $\pm$ 0.16 & -16.0670 $\pm$ 0.0013 & 12.8 & 170 & SB	\\
81 & 20419 & K8 & 9.79 & 1.183 $\pm$ 0.005 & 40.77 $\pm$ 0.20 & 39.46 $\pm$ 0.60 & 39.34 $\pm$ 0.16 & 33.7640 $\pm$ 0.0014 & 6.2 & 101 & SB	\\
82 & 20455 & G8III & 3.77 & 0.983 $\pm$ 0.010 & 39.65 $\pm$ 0.08 & 39.04 $\pm$ 0.60 & 38.92 $\pm$ 0.16 & 37.5875 $\pm$ 0.0004 & 8.9 & 213 & SB	\\
83 & 20661 & F7V & 6.44 & 0.509 $\pm$ 0.005 & 39.10 $\pm$ 0.50 & 39.48 $\pm$ 0.60 & 39.35 $\pm$ 0.16 & 41.7982 $\pm$ 0.0035 & 25.7 & 184 & SB	\\
84 & 20679 & K2V & 8.99 & 0.935 $\pm$ 0.005 & 37.00 $\pm$ 7.50 & 39.27 $\pm$ 0.60 & 39.15 $\pm$ 0.16 & 40.9931 $\pm$ 0.0011 & 10.3 & 119 & SB	\\
85 & 20686 & G5V & 8.07 & 0.680 $\pm$ 0.600 & 40.72 $\pm$ 0.47 & 39.17 $\pm$ 0.60 & 39.05 $\pm$ 0.16 & 37.8761 $\pm$ 0.0014 & 13.3 & 149 & SB	\\
86 & 20712 & F8V & 7.36 & 0.557 $\pm$ 0.013 & 38.77 $\pm$ 0.14 & 38.83 $\pm$ 0.60 & 38.71 $\pm$ 0.16 & 24.7244 $\pm$ 0.0013 & 8.5 & 97 & SB	\\
87 & 20751 & K5 & 9.45 & 1.033 $\pm$ 0.005 & 41.12 $\pm$ 0.20 & 39.93 $\pm$ 0.60 & 39.80 $\pm$ 0.17 & 45.8963 $\pm$ 0.0008 & 7.7 & 131 & SB	\\
88 & 20890 & G8V & 8.62 & 0.741 $\pm$ 0.014 & 39.91 $\pm$ 0.08 & 39.32 $\pm$ 0.60 & 39.20 $\pm$ 0.16 & 36.3041 $\pm$ 0.0007 & 7.8 & 96 & SB	\\
89 & 21112 & F9V & 7.78 & 0.540 $\pm$ 0.019 & 40.98 $\pm$ 0.31 & 40.22 $\pm$ 0.60 & 40.10 $\pm$ 0.17 & 40.7134 $\pm$ 0.0012 & 8.4 & 105 & SB	\\
90 & 21543 & G1V & 7.53 & 0.597 $\pm$ 0.012 & 42.00 $\pm$ 0.33 & 40.69 $\pm$ 0.60 & 40.57 $\pm$ 0.17 & 36.5396 $\pm$ 0.0018 & 11.6 & 112 & SB	\\
91 & 22203 & G5 & 8.30 & 0.665 $\pm$ 0.006 & 42.42 $\pm$ 0.71 & 41.44 $\pm$ 0.60 & 41.32 $\pm$ 0.17 & 41.3213 $\pm$ 0.0011 & 8.6 & 108 & SB	\\
92 & 22224 & K0 & 9.60 & 0.967 $\pm$ 0.005 & 40.32 $\pm$ 0.09 & 41.21 $\pm$ 0.60 & 41.09 $\pm$ 0.17 & 42.9301 $\pm$ 0.0015 & 7.6 & 58 & SB	\\
93 & 22524 & F8 & 7.29 & 0.536 $\pm$ 0.011 & 42.74 $\pm$ 0.17 & 41.72 $\pm$ 0.60 & 41.60 $\pm$ 0.17 & 35.2684 $\pm$ 0.0046 & 23.0 & 129 & SB	\\
94 & 23983 & A2m & 5.43 & 0.249 $\pm$ 0.010 & 44.16 $\pm$ 0.14 & 43.56 $\pm$ 0.60 & 43.45 $\pm$ 0.17 & 43.9119 $\pm$ 0.0036 & 17.2 & 113 & SB	\\
95 & 13834 & F5IV & 5.80 & 0.415 $\pm$ 0.009 & 28.10 $\pm$ 1.20 & 27.97 $\pm$ 0.58 & 27.83 $\pm$ 0.13 & 27.3858 $\pm$ 0.0066 & 35.6 & 136 & x	\\
96 & 18170 & F4V & 5.97 & 0.354 $\pm$ 0.017 & 35.00 $\pm$ 2.50 & 35.77 $\pm$ 0.59 & 35.63 $\pm$ 0.15 & 33.2054 $\pm$ 0.0394 & 82.8 & 119 & x	\\
97 & 18658 & F5V & 6.35 & 0.417 $\pm$ 0.012 & 39.10 $\pm$ 1.10 & 36.94 $\pm$ 0.59 & 36.81 $\pm$ 0.16 & 37.7031 $\pm$ 0.0231 & 79.8 & 133 & x	\\
98 & 18735 & F4V... & 5.89 & 0.319 $\pm$ 0.005 & 31.70 $\pm$ 1.10 & 36.58 $\pm$ 0.59 & 36.45 $\pm$ 0.15 & -13.1284 $\pm$ 0.0593 & 461.8 & 145 & x	\\
99 & 19261 & F3V & 6.02 & 0.397 $\pm$ 0.004 & 36.35 $\pm$ 0.26 & 37.65 $\pm$ 0.59 & 37.52 $\pm$ 0.16 & 36.3151 $\pm$ 0.0112 & 40.0 & 132 & x	\\
100 & 19504 & F6V & 6.61 & 0.427 $\pm$ 0.009 & 37.10 $\pm$ 0.30 & 37.66 $\pm$ 0.59 & 37.54 $\pm$ 0.16 & 37.1204 $\pm$ 0.0085 & 36.1 & 150 & x	\\
101 & 19591 & K0 & 9.38 & 1.090 $\pm$ 0.015 & 36.90 $\pm$ 0.26 & 37.03 $\pm$ 0.59 & 36.91 $\pm$ 0.15 & 37.0939 $\pm$ 0.0099 & 6.7 & 109 & x	\\
102 & 19789 & F5V & 7.05 & 0.424 $\pm$ 0.006 & 38.40 $\pm$ 1.20 & 37.50 $\pm$ 0.59 & 37.38 $\pm$ 0.15 & 29.9866 $\pm$ 0.0388 & 104.8 & 143 & x	\\
103 & 19877 & F5Vvar & 6.31 & 0.400 $\pm$ 0.015 & 36.40 $\pm$ 1.20 & 38.51 $\pm$ 0.60 & 38.39 $\pm$ 0.16 & 37.4085 $\pm$ 0.0309 & 81.2 & 134 & x	\\
104 & 20215 & F7V+... & 6.85 & 0.509 $\pm$ 0.015 & 39.21 $\pm$ 0.27 & 38.84 $\pm$ 0.60 & 38.72 $\pm$ 0.16 & 38.6342 $\pm$ 0.0028 & 19.4 & 151 & x	\\
105 & 20219 & F3V... & 5.58 & 0.283 $\pm$ 0.008 & 42.00 $\pm$ 2.50 & 39.06 $\pm$ 0.60 & 38.93 $\pm$ 0.16 & 3.9646 $\pm$ 0.0298 & 154.9 & 162 & x	\\
106 & 20349 & F5V & 6.79 & 0.434 $\pm$ 0.015 & 37.10 $\pm$ 1.20 & 38.43 $\pm$ 0.60 & 38.31 $\pm$ 0.16 & 36.3019 $\pm$ 0.0257 & 106.1 & 179 & x	\\
107 & 20350 & F6V & 6.80 & 0.441 $\pm$ 0.014 & 40.80 $\pm$ 2.40 & 38.80 $\pm$ 0.60 & 38.68 $\pm$ 0.16 & 37.8929 $\pm$ 0.0230 & 59.8 & 100 & x	\\
108 & 20400 & A3m & 5.72 & 0.315 $\pm$ 0.008 & 37.80 $\pm$ 2.30 & 39.27 $\pm$ 0.60 & 39.15 $\pm$ 0.16 & 65.7194 $\pm$ 0.0090 & 42.7 & 166 & x	\\
109 & 20491 & F5 & 7.18 & 0.462 $\pm$ 0.009 & 35.90 $\pm$ 0.50 & 38.04 $\pm$ 0.59 & 37.92 $\pm$ 0.15 & 37.4953 $\pm$ 0.0183 & 52.9 & 108 & x	\\
110 & 20542 & A7V & 4.80 & 0.154 $\pm$ 0.007 & 39.20 $\pm$ 1.20 & 39.17 $\pm$ 0.60 & 39.05 $\pm$ 0.16 & 37.8405 $\pm$ 0.0191 & 60.1 & 192 & x	\\
111 & 20567 & F6V & 6.96 & 0.450 $\pm$ 0.018 & 40.10 $\pm$ 0.60 & 39.24 $\pm$ 0.60 & 39.11 $\pm$ 0.16 & 38.6270 $\pm$ 0.0153 & 48.0 & 111 & x	\\
112 & 20842 & Am & 5.72 & 0.270 $\pm$ 0.015 & 37.50 $\pm$ 3.30 & 38.97 $\pm$ 0.60 & 38.85 $\pm$ 0.16 & 35.0437 $\pm$ 0.0275 & 119.7 & 185 & x	\\
113 & 20873 & F0IV & 5.90 & 0.325 $\pm$ 0.013 & 40.60 $\pm$ 0.30 & 39.86 $\pm$ 0.60 & 39.73 $\pm$ 0.16 & 36.6681 $\pm$ 0.0275 & 108.7 & 110 & x	\\
114 & 20901 & A7V & 5.02 & 0.215 $\pm$ 0.003 & 39.90 $\pm$ 4.10 & 40.02 $\pm$ 0.60 & 39.90 $\pm$ 0.17 & 33.7276 $\pm$ 0.0324 & 121.3 & 126 & x	\\
115 & 20948 & F6V & 6.90 & 0.451 $\pm$ 0.002 & 38.62 $\pm$ 0.24 & 39.65 $\pm$ 0.60 & 39.53 $\pm$ 0.16 & 38.9562 $\pm$ 0.0102 & 42.3 & 144 & x	\\
116 & 21008 & F6V & 7.09 & 0.470 $\pm$ 0.015 & 38.00 $\pm$ 2.50 & 39.46 $\pm$ 0.60 & 39.34 $\pm$ 0.16 & 39.1536 $\pm$ 0.0130 & 35.9 & 90 & x	\\
117 & 21029 & A6IV & 4.78 & 0.170 $\pm$ 0.001 & 41.00 $\pm$ 1.80 & 39.93 $\pm$ 0.60 & 39.81 $\pm$ 0.16 & 32.8012 $\pm$ 0.0430 & 88.0 & 144 & x	\\
118 & 21039 & Am & 5.47 & 0.258 $\pm$ 0.003 & 39.56 $\pm$ 0.23 & 39.99 $\pm$ 0.60 & 39.87 $\pm$ 0.16 & 39.1903 $\pm$ 0.0110 & 36.5 & 126 & x	\\
\bottomrule
 \end{tabular}
\end{table*}

\setcounter{table}{0}
\begin{table*}
 \caption{Continued.} 
 \centering
 \setlength{\tabcolsep}{0.54em}
 \begin{tabular}{cccccccc|cccc}
\toprule
\# & HIP & SpType & $V_{\rm mag}$ & $(B-V)$ & $RV_{\rm P98}^{\rm Spectr.}$ & $RV_{\rm M02}^{\rm Astrom.}$ & $RV_{\rm G17}^{\rm Astrom.}$ & $RV_{\rm HARPS}$ & FWHM & $S/N$ & Flag \\
 & & & (mag) & (mag) & (km~s$^{-1}$) & (km~s$^{-1}$) & (km~s$^{-1}$) & (km~s$^{-1}$) & (km~s$^{-1}$) & & \\
\midrule
119 & 21066 & F5 & 7.03 & 0.472 $\pm$ 0.013 & 41.35 $\pm$ 0.26 & 40.34 $\pm$ 0.60 & 40.22 $\pm$ 0.17 & 40.3800 $\pm$ 0.0118 & 40.2 & 115 & x	\\
120 & 21137 & F4V... & 6.01 & 0.338 $\pm$ 0.012 & 36.00 $\pm$ 2.50 & 40.09 $\pm$ 0.60 & 39.97 $\pm$ 0.16 & 13.7373 $\pm$ 0.0422 & 184.9 & 145 & x	\\
121 & 21152 & F5V & 6.37 & 0.420 $\pm$ 0.014 & 39.80 $\pm$ 1.00 & 40.46 $\pm$ 0.60 & 40.34 $\pm$ 0.17 & 39.5813 $\pm$ 0.0216 & 64.0 & 113 & x	\\
122 & 21267 & F5V & 6.62 & 0.429 $\pm$ 0.012 & 36.90 $\pm$ 0.90 & 40.49 $\pm$ 0.60 & 40.37 $\pm$ 0.17 & 39.5488 $\pm$ 0.0226 & 66.1 & 111 & x	\\
123 & 21459 & F5IV & 6.01 & 0.380 $\pm$ 0.600 & 43.30 $\pm$ 1.20 & 39.43 $\pm$ 0.60 & 39.32 $\pm$ 0.16 & 40.5994 $\pm$ 0.0392 & 107.3 & 136 & x	\\
124 & 21474 & F5V & 6.64 & 0.442 $\pm$ 0.018 & 33.70 $\pm$ 1.20 & 40.54 $\pm$ 0.60 & 40.42 $\pm$ 0.16 & 41.1699 $\pm$ 0.0199 & 65.6 & 120 & x	\\
125 & 21589 & A6V & 4.27 & 0.122 $\pm$ 0.005 & 44.70 $\pm$ 5.00 & 40.94 $\pm$ 0.60 & 40.82 $\pm$ 0.17 & 39.4782 $\pm$ 0.0547 & 112.9 & 168 & x	\\
126 & 21670 & A5m & 5.38 & 0.257 $\pm$ 0.009 & 36.30 $\pm$ 1.20 & 41.17 $\pm$ 0.60 & 41.05 $\pm$ 0.17 & 39.3510 $\pm$ 0.0326 & 77.4 & 114 & x	\\
127 & 22550 & F6V & 6.79 & 0.543 $\pm$ 0.013 & 42.44 $\pm$ 0.17 & 42.15 $\pm$ 0.60 & 42.03 $\pm$ 0.17 & 41.5021 $\pm$ 0.0016 & 10.7 & 123 & x	\\
128 & 22850 & F3IV & 6.36 & 0.292 $\pm$ 0.012 & 38.40 $\pm$ 2.00 & 41.61 $\pm$ 0.60 & 41.50 $\pm$ 0.17 & 24.9762 $\pm$ 0.0452 & 174.3 & 104 & x	\\
129 & 23214 & F5V & 6.75 & 0.450 $\pm$ 0.015 & 42.50 $\pm$ 1.50 & 42.44 $\pm$ 0.60 & 42.33 $\pm$ 0.17 & 42.2957 $\pm$ 0.0132 & 44.5 & 120 & x	\\
130 & 26382 & F0V & 5.53 & 0.237 $\pm$ 0.011 & 41.10 $\pm$ 1.20 & 44.49 $\pm$ 0.60 & 44.40 $\pm$ 0.17 & 28.8363 $\pm$ 0.0629 & 142.5 & 88 & x	\\
131 & 28356 & F5IV & 7.78 & 0.461 $\pm$ 0.015 & 45.00 $\pm$ 2.50 & 45.83 $\pm$ 0.59 & 45.74 $\pm$ 0.18 & 60.0714 $\pm$ 0.0212 & 63.1 & 142 & x	\\
\bottomrule
 \end{tabular}
\end{table*}

\subsection{Spectroscopic observations and data reduction}\label{sec:data}

All spectra were acquired with the HARPS 
\citep[High Accuracy Radial velocity Planet Searcher,][]{Mayor03} 
high-resolution spectrograph (R~=~115\,000), fed by the 3.6~m telescope 
in La Silla, Chile. The spectral range covers from 3800 to 6900~\AA, with a 
small gap between 5300--5330~\AA~because of the arrangement of the CCD mosaic. 
 All the spectra were reprocessed by the last version of the HARPS pipeline (Data Reduction Software version 3.5). 
All stars have been observed once, in some case observations have been repeated, if the first spectrum casted doubts on the quality of the data. 
The S/N ratio was always exceeding 50, that implies a photon noise precision of 2~m~s$^{-1}$ or better for a slow rotating G star; 
as explained above, photon noise is not limiting the accuracy of 
the measurements. The list of the observed stars and the measured RV
is given in Table~\ref{table:main}. 

All the spectra were cross correlated with the same digital G2 mask, based on the observed solar spectrum and optimised for the
HARPS spectrograph. Being the mask based on an observed solar spectrum, we do not expect a large zero point correction,
in fact the G2 HARPS mask has been 
recently calibrated by \citet{Lanza16} using solar system bodies, who found a zero point shift of $\sim$100~m~s$^{-1}$. 
We will come back on this point in Sect.~\ref{sec:m1}.

All the cross-correlation functions (CCFs) and their fits have been checked visually, and a number of stars show lower quality CCFs,
either because they present a slope in the CCF continuum, or because multiple peaks are observed, or because the CCF has a non gaussian shape. 
This is not too surprising, when considering that the original astrometric sample included binaries, as well as stars spanning a large range of spectral types.
Hot stars have many fewer lines than the G2 mask and often much broadened by high stellar rotation. 
Cool stars have many more lines, and therefore present a richer and more complex spectrum. Nevertheless we 
consider the use of the same, well calibrated mask for all stars an enormous advantage to perform our investigation.

\subsection{Astrometric radial velocities}\label{sec:astromrvs}

Astrometric RVs for our initial sample were taken from M02, choosing the ``Hipparcos'' solution, 
that is the one recommended by the authors. 
M02 computed the RVs from their Eqs. (1)--(3).
The critical input to these equations is the space velocity of the cluster,
which is then combined with the right ascension $\alpha$ and declination $\delta$ of the cluster members.
These parameters are enough to estimate astrometric RVs from a simple geometrical model assuming that all the cluster
members are moving with the same space velocity of the cluster, with no acceleration.
Possible accelerations originated, for example,
from cluster rotation or expansion are not considered and may introduce biases (see Sect.~\ref{sec:asymmetry}).
The expected error in the astrometric RV from M02 for each star is of the order of 600~m~s$^{-1}$.
 Since the astrometric radial velocity is the the space motion of the cluster projected onto the line of sight for each star,
 the astrometric radial velocity error is dominated by the error in the cluster velocity vector, and is highly correlated from star to star. 

As far as {\em Gaia}, G17 provide a more refined space velocity of the cluster in the 3$\times$2 matrix given by their Eq.~(4).
The first column of the matrix provides the space velocities computed from astrometric information only.
The second column uses astrometric and spectroscopic data, so the resulting velocity is affected
by the phenomena influencing the spectroscopic RVs discussed in this work. We therefore computed our {\em Gaia} based astrometric RVs
from the pure astrometric solution of the space velocity:
$V_{x} = -6.03 \pm 0.08$~km~s$^{-1}$, $V_{y} = 45.56 \pm 0.18$~km~s$^{-1}$, and $V_{z} = 5.57 \pm 0.06$~km~s$^{-1}$.
The {\em Gaia} astrometric RVs are then calculated by simply using Eq.~(2) of M02, which is given by:
\begin{equation}
RV_{\rm G17}^{\rm Astrom.} = V_x\cos\delta\cos\alpha + V_y\cos\delta\sin\alpha + V_z\sin\delta.
\label{eq:rv}
\end{equation}
No parallax is needed and the errors induced by the uncertainty in the ($\alpha$, $\delta$) coordinates is negligible.
The typical error associated to the astrometric RVs (G17 RVs hereafter) is around~160~m~s$^{-1}$, almost one-forth of the M02 errors. 
Additional astrometric data are provided in Appendix~\ref{a:addastro}.

\subsection{Mass to radius ratio}\label{sec:physparam}

Since gravitational redshift depends on the stellar mass to radius ratio, $M/R$,
this ratio was computed from the $(B-V)$ color by using theoretical isochrones,
obtained from the CMD\footnote{\url{http://stev.oapd.inaf.it/cgi-bin/cmd}} Web Interface \citep[e.g.,][]{Bressan12,Tang14,Chen14,Chen15}.
The used isochrone has log(age/yr) = 8.8 with initial hydrogen and helium compositions 
(based on the XYZ Calculator\footnote{\url{http://astro.wsu.edu/cgi-bin/XYZ.pl}} web interface with 
the \citealt{Grevesse98} values) $Z = 0.024$ and $Y = 0.292$. Polynomial fits were obtained for the main sequence and 
red giant branch separately for the theoretical $(B-V)$ versus $M/R$ functions, from which photometric
 $M/R$ values were estimated for all the sample stars.
Uncertainties were estimated with a Monte Carlo approach by applying fluctuations 
to the observed $(B-V)$ values within their errors.

\section{Spectroscopic vs. astrometric RV}\label{sec:analysis}

\begin{figure}
\centering
\includegraphics[width=0.45\textwidth]{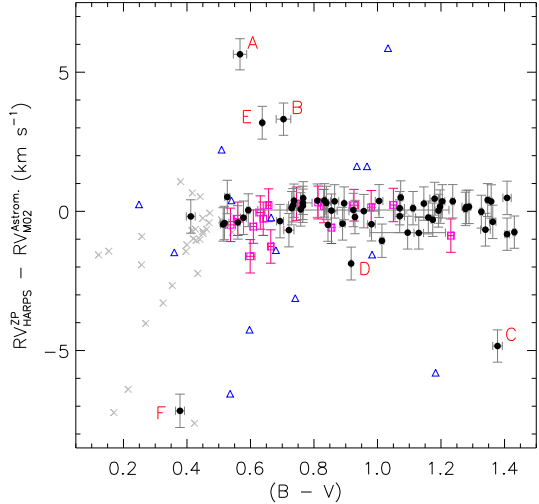}
\caption{Difference $\Delta RV_{\rm M02}^{\rm ZP}$ between the spectroscopic RV corrected for zero point ($RV_{\rm HARPS}^{\rm ZP}$)
 and the astrometric RV ($RV_{\rm M02}^{\rm Astrom.}$) from M02 vs. $B-V$ for all the stars 
 observed with HARPS. $RV_{\rm HARPS}^{\rm ZP}$ refers the HARPS RV measurements corrected for the zero point of 102~m~s$^{-1}$ (see text).
 Its range is truncated for a close view of the data, and all stars beyond the bounds are binaries or have bad quality CCFs.
 Different symbols refer to quality classification: filled black circles for the selected data; 
 grey crosses for the bad quality CCFs; blue triangles for spectroscopic binary stars (identified from the literature 
 or from visual inspection of the CCFs); and purple squares for known BY~Dra stars.
 Targets A--F are potential outliers described in Sect.~\ref{sec:analysis}.
 The magnitude $V$ and the $(B-V)$ color are from P98 and all stars shown fulfill the membership criterion of this work.}
\label{fig:bvvsdelrv}
\end{figure}

M02 sample was carefully selected for Hyades members that share the space velocity of the cluster with a low velocity spread.
Out of the 168 stars in the M02 list, 131 were observed with HARPS. 
They constitute our initial sample since they have both, HARPS spectroscopic RV and M02 astrometric RV.
After cleaning the M02 sample, we computed the G17 RVs for the finally selected stars.

 When comparing spectroscopic and astrometric RV, we would expect, in the ideal case, a gaussian distribution. The centre representing the agreement between the two measurements
 and the dispersion of the distribution showing the velocity dispersion of the cluster (plus other sources of measurement noise). The cluster velocity dispersion should in fact dominate the
 width of the difference, because it will affect the spectroscopic RVs, but not the astrometric ones. The Hyades velocity dispersion is evaluated around 300--320~m~s$^{-1}$ (P98; \citealt{Reino18}).

The difference $\Delta RV_{\rm M02}^{\rm ZP}$ between the spectroscopic RV corrected for zero point
(see Sect.~\ref{sec:m1}), namely $RV_{\rm HARPS}^{\rm ZP}$, and the astrometric RV from M02, namely $RV_{\rm M02}^{\rm Astrom.}$, for the 131 sample stars 
is shown in Fig.~\ref{fig:bvvsdelrv}. Different symbols indicate the quality of the fits, the binarity or stars with 
variable RV. Note that this Figure is limited to a $\sim\pm$~7~km~s$^{-1}$ range and several stars lay beyond these limits.
This figure is equivalent and very similar to Fig.~4 of M02. 
The distribution of $\Delta RV_{\rm M02}^{\rm ZP}$ has an 
average value of $-2.0$~km~s$^{-1}$, and a very broad standard deviation ($\sigma$) of 
8.5~km~s$^{-1}$. It is noticeable that many 
stars have a $\Delta RV_{\rm M02}^{\rm ZP}$ of several km~s$^{-1}$, in large excess of the expected measurement errors and that the 
bulk of the sample is centred in a well defined sequence close to zero.

In order to have a more useful comparison, we cleaned the original sample from all the spectroscopic RV variables, such as
known spectroscopic binaries and Delta Scuti stars. 
Several studies investigated binaries in the Hyades, in particular \citet{Griffin98,Griffin12} summarise the results of many 
years of spectroscopic monitoring, identifying a number of spectroscopic binaries, which were first eliminated from the sample. 
Spectroscopic RVs of Delta Scuti stars are also known to vary by several km~s$^{-1}$ \citep{Poretti01,Poretti09} 
and these variables were also eliminated from the comparison. 

In addition, the CCF peaks of some of the stars observed with HARPS are double, indicating that they are spectroscopic 
binaries. These stars (labelled as ``SB'' in the quality flag of table~\ref{table:main}) were also removed. 
Finally a number of stars simply show rather bad CCFs. 
All of them are hot stars and the bad CCF profile is likely caused by the mismatch between the G2 mask 
and their early spectral types. We removed them as well from the final sample, although their exclusion 
or inclusion does not change substantially our main results. 
The likely single stars with bona fide CCFs (61 stars with flags 0 or 1 in Table~\ref{table:main}) 
and the BY~Dra variables (16 stars with ``BY'' flag) form our best sample, for a total of 77 stars.
Not surprisingly, they concentrate around the solar-type region of the C-M diagram and cooler. 

The mean difference $\Delta RV_{\rm M02}^{\rm ZP}$ for this sample is 
of $-101$~m~s$^{-1}$ with a rather large spread, with $\sigma$ of 1.4~km~s$^{-1}$. 
In fact, in spite all the known spectroscopic RV variables have been eliminated from the sample, a few stars 
show large spectroscopic -- astrometric differences (larger than 2~km~s$^{-1}$) (cf. Fig.~\ref{fig:bvvsdelrv}); we searched in 
literature for more information on these stars and for all of them but one there is evidence for binarity or doubtful
 membership. Only for star (F) HIP 20614 we do not find any indication of binarity, 
 however we notice that for this star the HARPS RV and the RV published by P98 
 differ by 5~km~s$^{-1}$, indicating that its RV varies in the long period.
 We provide below a short summary of the stars with large spectroscopic and astrometric differences 
 which have been discarded from our best sample. 
\begin{itemize}
\item {\em (A) HIP 10672}. This star has a metallicity of $[Fe/H]= -0.12$~dex according to 
\citet{Paulson04}. In addition, the HARPS RV deviates from the published RV values.
\item {\em (B) HIP 13600}. Chemical abundances deviate from those of the Hyades \citep{Paulson03}. Therefore, this object is likely a nonmember.
\item {\em (C) HIP 16548 and (D) 16908}. Both stars in the paper by \citet{Guenther05} have 
low luminosity companion, about 3 mag fainter. One has a separation of 0.7 arc sec, one of 2.7.
\item {\em (E) HIP 19365}. Given as potential binary in Simbad. 
\item {\em (F) HIP 20614}. Nothing special identified in different works (\citealt{Hartkopf01}; 
\citealt{BohmVitense02}; P98), and no companion detected by speckle interferometry \citep{Mason09}. 
Given the difference with previously published RV, we argue that also this star is a long period spectroscopic binary.
\end{itemize}

\begin{figure}
\centering
\includegraphics[width=0.45\textwidth]{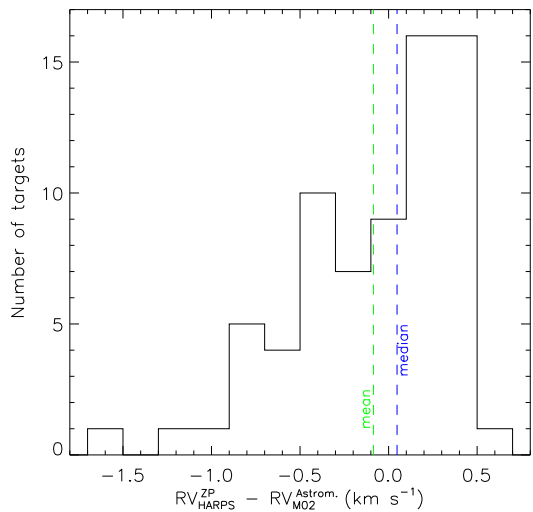}
 \caption{Distribution of the spectroscopic -- astrometric RV differences (corrected for zero point), $\Delta RV_{\rm M02}^{\rm ZP}$, for the M02 astrometric solution for the best quality subsample (of 71 stars with flags 0, 1, and ``BY'', and outliers excluded). The mean and median values of the distribution are illustrated by the green and blue dashed lines, respectively.}
\label{fig:histdelrv}
\end{figure}

After removing these outliers, 71
objects are left. The average difference between spectroscopic (corrected for zero point) and astrometric RVs 
as measured from HARPS and Hipparcos (M02), namely $\Delta RV_{\rm M02}^{\rm ZP}$, is of $-85$~m~s$^{-1}$, with 
a $\sigma$ of 464~m~s$^{-1}$.
The distribution of $\Delta RV_{\rm M02}^{\rm ZP}$, displayed in Fig.~\ref{fig:histdelrv}, is non gaussian, and the median is of +48~m~s$^{-1}$.
It is intriguing that this distribution is not gaussian, but skewed and possibly double-peaked.
We discuss in Sect.~\ref{sec:asymmetry} different possibilities for the origin of such asymmetry.

A cross-match between our final sample of 71 objects and the {\em Gaia} Archive\footnote{\url{https://gea.esac.esa.int/archive/}} results in a subsample of 65 objects with new astrometric data, including parallax. The astrometric RV can also be computed for the missing 6 objects by using equatorial coordinates from another
catalogue in Eq.~(2) of M02. We used the Hipparcos coordinates, which differ by typically $\sim 3$~arcsec from the {\em Gaia} values.
This difference is negligible in the astrometric RV calculation. 
The distribution of the spectroscopic (from HARPS, corrected for zero point) minus astrometric RVs from G17, namely $\Delta RV_{\rm G17}^{\rm ZP}$,
is similar to the $\Delta RV_{\rm M02}^{\rm ZP}$ distribution displayed in Fig.~\ref{fig:histdelrv}, just shifted by an offset.
Therefore, the asymmetry of the $\Delta RV_{\rm M02}^{\rm ZP}$ distribution is also observed in the $\Delta RV_{\rm G17}^{\rm ZP}$ distribution.
The $\sigma$ value of the $\Delta RV_{\rm G17}^{\rm ZP}$ distribution is 462~m~s$^{-1}$, similar to the $\sigma$ of $\Delta RV_{\rm M02}^{\rm ZP}$.

\subsection{Accuracy of spectroscopic RV}\label{sec:m1}

Most spectroscopic RV applications (e.g. clusters membership, search for binaries or exo-planets)
are interested in obtaining the highest measurements precision. The present study, that compares results obtained with
different techniques, requires as well a high accuracy. 

The zero point of the G2 mask of HARPS has been measured to an accuracy of a few m~s$^{-1}$, by comparing spectroscopic RV and
true velocities of solar system bodies, by \citet{Lanza16}, who find a
(relativistic corrected) shift that varies between 
98 and 102~m~s$^{-1}$, correlating with the solar activity cycle. 
Our observations were taken in December 2014, in a period not included in the Lanza et al. study; however, 
since in the last years of the Lanza study the zero point of the HARPS G mask 
has been rather stable with time, we consider a zero point of 102~m~s$^{-1}$ appropriate for our observations.
This zero point correction must therefore be added to all the data of Table~\ref{table:main}.
We shall keep in mind that an additional, systematic uncertainty of $\pm$2~m~s$^{-1}$ is associated to our HARPS
observations when performing the zero point correction because of its uncertainty.
For most spectra the formal error to the fit of the HARPS CCF is smaller than 
this zero point uncertainty. We also note that after the change of HARPS 
fibres to octagonal ones, the zero point of the G2 HARPS mask has moved by 
further $\sim$12~m~s$^{-1}$ \citep{LoCurto15,Molaro16}. 

Other aspects of stellar physics affect the spectroscopic 
RV accuracy \citep[see][]{Lindegren03}.
Stellar activity, for instance, may contribute with two effects.
 The first effect is that rotating inhomogeneities on stellar surface or long-term activity cycles distort the spectral line profiles,
 producing shifts in the observed RVs of individual targets. However, the shifts are modulated with time, and by observing many stars we shall cover random rotation or activity-cycle phases. The net result is that 
 we do not expect a systematic shift from this effect; rather an extra jitter in the spectroscopic radial velocity distribution, that will add to the cluster velocity dispersion
 (but it is almost one order of magnitude smaller).
It is not easy to evaluate the RV variability introduced by activity, but several authors estimate that jitter to less than 40~m~s$^{-1}$ for the Hyades \citep{Paulson04,Saar97}. The second effect of activity
 may, instead, introduce a systematic shift, because all the Hyades cluster stars are enhanced in activity with respect to the Sun, and all our zero point corrections, either empirical or theoretical, are suitable for a quiet star.
These (and other) effects are separately discussed in Sect.~\ref{sec:rvastrovsspec}.

\subsection{Accuracy of astrometric RVs\label{sec:astrometric}}

The difference between the M02 and G17 RVs is dominated by the offset caused by the different 
cluster space velocity found by the two authors.
Being the astrometric radial velocities the projection of the coordinates vector on the space velocity vector, the use of a
different space velocity translates into an offset, which is of $\left< RV_{\rm G17} - RV_{\rm M02} \right> = -123$~m~s$^{-1}$,
with a narrow $\sigma$ of 8~m~s$^{-1}$ for our case.

The formal errors for the astrometric RV computed in M02 are given by \citet{Lindegren00} 
and are of the order of 600~m~s$^{-1}$.
We must be careful on the interpretation of these astrometric uncertainties, because 
M02 mention that their alternative solution, based of Tycho values, has 
an offset of $-900$~m~s$^{-1}$ with respect to the 
Hipparcos one. Similarly, \citet{vanLeeuwen09} proposes a space motion for the Hyades
that is also about 1~km~s$^{-1}$ lower than the M02 one. 
The \citet{deBruijne02} solution as well as most solutions quoted in his paper and based on Hipparcos results, are 
compatible with M02. In their Table~1 \citet{deBruijne01} compare the Hyades space motions 
derived in several studies, and the majority predicts an astrometric RV for the cluster between 
39.44 and 39.51~km~s$^{-1}$ (70~m~s$^{-1}$ span); 
only the P98 and the Van Leeuwen solutions (plus the Tycho 2 solution of M02) form a distinct group, 
all providing a cluster RV about 1~km~s$^{-1}$ smaller.
These discrepancies should
induce some caution in (over-) interpreting the results.
The comparison with the {\em Gaia} solution is however reassuring, because the M02 Hipparcos solution and the {\em Gaia} G17 RVs agree to $\sim$120~m~s$^{-1}$.

\subsection{The cluster rotation}\label{sec:asymmetry}

\begin{figure}
\centering
\includegraphics[width=0.45\textwidth]{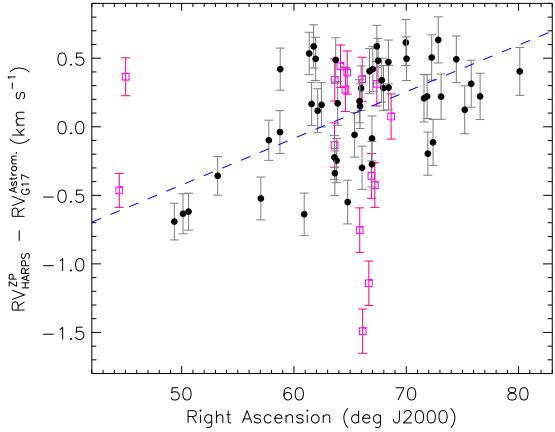}
 \caption{Spectroscopic -- astrometric G17 RV differences (corrected for zero point), $\Delta RV_{\rm G17}^{\rm ZP}$, vs. right 
 ascension ($\alpha$) for our stars. Dashed blue line is a linear fit (computed by excluding the BY~Dra variables; see text) illustrating the correlation between these quantities. Target symbols and colours follow the same description as in Fig.~\ref{fig:bvvsdelrv}.}
\label{fig:ravsdelv}
\end{figure}

In order to understand the origin of the asymmetry and the overall shape of the $\Delta RV_{\rm M02}^{\rm ZP}$ and $\Delta RV_{\rm G17}^{\rm ZP}$ distributions,
we have looked at the dependence of $\Delta RV_{\rm G17}^{\rm ZP}$ on several parameters. The same analysis can be repeated with the M02 data,
and provides same results, but with higher uncertainties.
Most significant is the correlation between $\Delta RV_{\rm G17}^{\rm ZP}$ and the 
stars' right ascension ($\alpha$), as shown in Fig.~\ref{fig:ravsdelv}.
The Pearson correlation coefficient of this plot is 0.35 for all the 71 targets and 0.56 if we remove the BY~Dra variables (purple squares),
which show a larger noise than the rest of the sample.
A linear fit without considering the BY~Dra variables gives that $\Delta RV_{\rm G17}^{\rm ZP}$ 
increases with $\alpha$ by $\sim$34~m/s/deg: $ \Delta RV_{\rm G17}^{\rm ZP} = (0.0340 \pm 0.0032) \alpha - (2.12 \pm 0.21)$~km~s$^{-1}$ (dashed blue line in the figure). 
A much weaker correlation for the declination
($\delta$) is also present in the data, but this is because $\alpha$ and $\delta$ are not fully independent. When $\Delta RV_{\rm G17}^{\rm ZP}$
is corrected for the $\alpha$ dependence, no correlation of $\Delta RV_{\rm G17}^{\rm ZP}$ with $\delta$ is left.

We tested and excluded two hypotheses that could explain such a dependence of $\Delta RV_{\rm G17}^{\rm ZP}$ with the target coordinates: inaccurate velocity vector and cluster expansion.
The tests consisted of several trials of a simple Hyades model based on the derived parameters for its 3D position and space velocity.
For the first hypothesis, we let the the velocity vector components free, in order to minimize the correlation between $\Delta RV_{\rm G17}^{\rm ZP}$ and $\alpha$. Because the changes in velocity should be greater than 1 km/s for such a minimization,
we consider this cause unlikely. For the cluster expansion, we verified that adding a radial expansion velocity for the cluster does produce an asymmetry in the RV distribution. However, that velocity should be much larger than the upper limit
 of 70 m~s$^{-1}$ estimated for the Hyades (see Sect.~\ref{sec:clexpantion}) to reproduce the observed distribution of $\Delta RV_{\rm G17}^{\rm ZP}$. We therefore also discarded this possibility.

A natural way of producing the $\Delta RV_{\rm G17}^{\rm ZP}$ dependence on $\alpha$
rotation of the cluster.
Intuitively, a cluster rotation should naturally produce a dependence of RV with star position, possibly a double peak in the RV distribution,
and  a skewness if the observed stars are not distributed homogeneously with respect to the cluster center.
We can convert  the dependence on $\alpha$ into velocity  gradient by assuming a distance of the cluster center at  46.09~parsecs
\citep{Reino18}, obtaining a rotation of 42.27~m~s$^{-1}$~pc$^{-1}$.

We note  that \citet{Vereshchagin13a} found indications of cluster rotation, with a gradient similar to ours,
but with low statistical significance. More recently, Reino et al. (2018), based on GAIA DR2,
concluded that the astrometric data alone do not support a significant rotation of the cluster.
Our results provide evidence that the Hyades cluster rotates and therefore
the comparison between spectroscopic and astrometric RV requires a correction for cluster rotation in the astrometric model.

\section{Phenomena affecting spectroscopic and astrometric RV}\label{sec:rvastrovsspec}

\subsection{Gravitational redshift}\label{sec:gravrsh}

Gravitational redshift has been predicted by \citet{Einstein17}. The GR effect scales with the solar mass and radius according to
\begin{equation}\label{eq:gr}
 \Delta RV_{GR} = 636.5 \times (M/R) - 3.0 \rm{~~m~s}^{-1}
\end{equation}
(for $M$ and $R$ given in Solar units), where the first term is the redshift as the light emitted 
from the solar photosphere was observed from infinity and the second term is the correction 
by placing the observer at the mean Earth distance from the Sun \citep[e.g.,][]{Lindegren03}.
Since the ratio $M/R$ may change easily by a factor 10 
in an open cluster when comparing main sequence and giant stars (the effect is much larger for white dwarfs, but they 
are not studied here), this is a very relevant effect, because spectroscopic RV are affected by GR, 
while astrometric ones are not. 
The measurement of GR in the Sun is not trivial, and several attempts have shown a general 
 agreement with the Einstein prediction, but with a limited accuracy \citep{Beckers77}. \citet{Pasquini11}
 studied spectroscopic RVs for many stars with different masses and radii in 
 the open cluster M67, but did not find evidence for GR. 
 They used 3D atmospheric models to simulate lines of different equivalent widths in giants and dwarfs, 
 finding that the lines in the dwarfs are blue-shifted with respect to giant stars by an amount that largely compensates 
 the differential GR. M67 is of solar chemical composition \citet{Randich00}, 
 while the Hyades are metal rich \citep{DutraFerreira16}. The instrument and mask used by \citet{Pasquini11}
 were also different from the one adopted here.

We estimated GR by computing the $M/R$ for each star, as described in Sect.~\ref{sec:physparam}.
The GR and $M/R$ values are given in Table~\ref{table:shifts} together with other parameters that are described below.
An analysis of the spectroscopic -- astrometric RV differences allows us to measure the GR effect from our data. This analysis is presented in Sect.~\ref{sec:maineffects}.

\subsection{Convective line shifts\label{sec:convshift}}

Convective blue shifts in stars are expected to be a major contributor to spectroscopic RVs, at a level 
comparable with GR, and being of opposite sign to GR, they will partially compensate for this effect.

Using a recent library of synthetic spectra based on hydrodynamical model
atmospheres \citep{Ludwig17} we estimated convective line shifts for 75 stars
of our ``best'' sample (two stars were too hot to be covered by the model
calculations). This was done by cross-correlating the synthetic spectra with
the G2-mask applied in the \HARPS\ data reduction pipeline
\citep{2002A&A...388..632P}. Effects of the finite spectral resolution of
\HARPS\ and stellar rotation were taken into account (see
Appendix~\ref{a:vconv} for details). The predicted shifts are listed in
Table~\ref{table:shifts}.

We emphasize that the listed values are the results of the synthetic profiles convolved 
with the \HARPS\ mask, so, while they are appropriate for the present work and for 
\HARPS\ observations in general, they should not be taken as absolute values of
the convective line shift in the particular star.

As sanity check we compared predicted shifts to the velocity zero point of
+102\pun{m s$^{-1}$} measured for the solar spectrum with \HARPS\ by Lanza et
al. (2016).
We used reduced spectra provided at the ESO\footnote{\url{https://www.eso.org/sci/facilities/lasilla/instruments/harps/inst/monitoring/sun.html}} webpage (Collection of HARPS solar spectra)
of Ceres, Ganymede, the Moon, and a daylight solar spectrum
to estimate the full width at half maximum (FWHM) of the cross-correlation as observed by \HARPS.
Our calculations gave an FWHM of
$7.2 \pm 0.1$~km~s$^{-1}$ for Ceres, Ganymede, and the Sun and $7.4 \pm 0.1$~km~s$^{-1}$ for the Moon.
For the 7.2~km~s$^{-1}$ width, we get a predicted shift for the solar spectrum of
$-490.2$~m~s$^{-1}$.
Assuming a solar gravitational redshift of $+633.5$\pun{m
s$^{-1}$} this results in a total zero point correction of $+143.3$\pun{m s$^{-1}$},
 to be compared to the value of Lanza and collaborators of $+102$\pun{m s$^{-1}$}. 
The difference of $41.3$\pun{m s$^{-1}$} is within the numerical accuracy we expect for the
theoretical spectra reflecting the underlying kinematics of flows in the 3D
model atmospheres. Further possible sources of uncertainties are: wavelengths
errors in the line list applied in the spectral synthesis, neglect of
departures from local thermodynamic equilibrium in the spectral synthesis, and
shortcomings of our approximate cross-correlation procedure when comparing to
pipeline results. However, in view of the modest velocity offset found, we consider the
correspondence satisfactory.

Having checked that the models provide sensible results for the Sun, we have applied the 
theoretical corrections for the two dominant effects, GR and convection, to all the stars, 
and the $\Delta RV_{\rm M02}^{GR+C}$ and $\Delta RV_{\rm G17}^{GR+C}$ values (the $GR+C$ superscript means corrected for GR and convection) so found are tabulated in Table~\ref{table:shifts}.
An analysis of these corrections is presented in Sect.~\ref{sec:maineffects}.

\begin{table*}
 \caption{Gravitational and convective shifts.}
 \label{table:shifts}
 {\centering
 \begin{tabular}{ccc|ccc|ccc|ccc}
\toprule
\# & HIP & $M/R$ & $GR$ & $C$ & $GR+C$ & $\Delta RV_{\rm M02}$ & $\Delta RV_{\rm M02}^{GR}$ & $\Delta RV_{\rm M02}^{GR+C}$ & $\Delta RV_{\rm G17}$ & $\Delta RV_{\rm G17}^{GR}$ & $\Delta RV_{\rm G17}^{GR+C}$ \\
 & & & (m s$^{-1}$) & (m s$^{-1}$) & (m s$^{-1}$) & (m s$^{-1}$) & (m s$^{-1}$) & (m s$^{-1}$) & (m s$^{-1}$) & (m s$^{-1}$) & (m s$^{-1}$) \\
\midrule
2 & 16529 & 1.130 $\pm$ 0.041 & 716 & -340 & 377 & -383 & -1100 & -760 & -255 & -972 & -632 \\
4 & 18327 & 1.135 $\pm$ 0.046 & 720 & -318 & 402 & 391 & -329 & -11 & 523 & -197 & 121 \\
5 & 19098 & 1.135 $\pm$ 0.077 & 719 & -327 & 392 & -341 & -1061 & -733 & -214 & -934 & -606 \\
6 & 19148 & 1.061 $\pm$ 0.070 & 672 & -563 & 109 & 142 & -530 & 33 & 269 & -404 & 160 \\
8 & 19781 & 1.097 $\pm$ 0.061 & 696 & -449 & 246 & -249 & -945 & -495 & -121 & -816 & -367 \\
9 & 19796 & 1.028 $\pm$ 0.052 & 651 & -573 & 79 & -368 & -1020 & -447 & -236 & -888 & -315 \\
10 & 20146 & 1.106 $\pm$ 0.048 & 701 & -451 & 250 & -574 & -1275 & -823 & -447 & -1147 & -696 \\
11 & 20205 & 0.248 $\pm$ 0.015 & 155 & -566 & -412 & -362 & -516 & 50 & -241 & -396 & 170 \\
12 & 20480 & 1.115 $\pm$ 0.050 & 707 & -389 & 318 & 171 & -535 & -146 & 289 & -418 & -29 \\
13 & 20492 & 1.132 $\pm$ 0.064 & 717 & -340 & 378 & 128 & -590 & -250 & 252 & -465 & -126 \\
14 & 20557 & 1.030 $\pm$ 0.041 & 652 & -602 & 51 & -316 & -969 & -367 & -196 & -849 & -247 \\
16 & 20826 & 1.047 $\pm$ 0.039 & 663 & -615 & 48 & -296 & -960 & -345 & -171 & -834 & -219 \\
17 & 20850 & 1.130 $\pm$ 0.056 & 716 & -345 & 371 & 404 & -312 & 33 & 523 & -193 & 152 \\
18 & 20889 & 0.220 $\pm$ 0.013 & 137 & -562 & -425 & -955 & -1092 & -530 & -838 & -975 & -413 \\
19 & 20949 & 1.117 $\pm$ 0.051 & 708 & -392 & 315 & 577 & -131 & 262 & 689 & -18 & 374 \\
20 & 20978 & 1.133 $\pm$ 0.038 & 718 & -313 & 406 & 461 & -257 & 55 & 584 & -134 & 179 \\
21 & 21099 & 1.109 $\pm$ 0.064 & 703 & -413 & 290 & 321 & -382 & 30 & 442 & -261 & 151 \\
22 & 21741 & 1.125 $\pm$ 0.059 & 713 & -391 & 322 & 486 & -227 & 164 & 599 & -115 & 277 \\
23 & 22380 & 1.129 $\pm$ 0.043 & 716 & -355 & 360 & 496 & -220 & 136 & 608 & -108 & 247 \\
24 & 22422 & 1.054 $\pm$ 0.053 & 668 & -623 & 45 & -127 & -795 & -172 & -11 & -680 & -56 \\
25 & 22566 & 1.034 $\pm$ 0.048 & 655 & -617 & 38 & 618 & -37 & 580 & 736 & 81 & 698 \\
26 & 23069 & 1.110 $\pm$ 0.060 & 703 & -396 & 308 & 484 & -219 & 177 & 595 & -109 & 287 \\
27 & 23498 & 1.116 $\pm$ 0.063 & 708 & -408 & 299 & 302 & -405 & 3 & 417 & -291 & 117 \\
28 & 23750 & 1.108 $\pm$ 0.078 & 702 & -438 & 264 & 215 & -487 & -49 & 324 & -378 & 60 \\
29 & 24923 & 1.116 $\pm$ 0.063 & 708 & -410 & 297 & 406 & -301 & 109 & 506 & -201 & 209 \\
31 & 15300 & 1.016 $\pm$ 0.077 & 644 & -289 & 355 & -719 & -1362 & -1074 & -589 & -1233 & -944 \\
32 & 15563 & 1.119 $\pm$ 0.033 & 709 & -236 & 473 & -674 & -1383 & -1147 & -532 & -1241 & -1005 \\
33 & 15720 & 1.003 $\pm$ 0.080 & 636 & -286 & 350 & -645 & -1281 & -995 & -516 & -1152 & -866 \\
35 & 17766 & 1.049 $\pm$ 0.050 & 665 & -243 & 422 & -557 & -1222 & -980 & -421 & -1086 & -843 \\
36 & 18018 & 1.112 $\pm$ 0.087 & 705 & -229 & 476 & -118 & -823 & -593 & 4 & -701 & -471 \\
37 & 18322 & 1.129 $\pm$ 0.056 & 716 & -236 & 480 & -75 & -791 & -554 & 63 & -652 & -416 \\
38 & 18946 & 1.126 $\pm$ 0.079 & 713 & -236 & 478 & -663 & -1377 & -1141 & -535 & -1249 & -1013 \\
39 & 19082 & 1.046 $\pm$ 0.118 & 663 & -230 & 433 & 510 & -153 & 77 & 637 & -26 & 204 \\
40 & 19207 & 1.107 $\pm$ 0.056 & 702 & -234 & 467 & 556 & -146 & 89 & 689 & -13 & 221 \\
41 & 19263 & 1.136 $\pm$ 0.065 & 720 & -260 & 460 & 473 & -247 & 13 & 598 & -122 & 138 \\
42 & 19316 & 1.055 $\pm$ 0.144 & 669 & -273 & 396 & 89 & -579 & -307 & 220 & -449 & -176 \\
43 & 19441 & 1.104 $\pm$ 0.062 & 700 & -231 & 469 & 126 & -574 & -343 & 262 & -437 & -206 \\
44 & 19808 & 1.100 $\pm$ 0.065 & 697 & -231 & 466 & 461 & -236 & -5 & 590 & -108 & 124 \\
45 & 19834 & 1.039 $\pm$ 0.075 & 658 & -351 & 307 & -275 & -934 & -582 & -144 & -802 & -451 \\
46 & 19862 & 1.137 $\pm$ 0.227 & 721 & -286 & 434 & 150 & -571 & -284 & 273 & -448 & -161 \\
47 & 20357 & 0.983 $\pm$ 0.046 & 622 & -522 & 101 & -82 & -704 & -182 & 44 & -579 & -57 \\
48 & 20527 & 1.072 $\pm$ 0.080 & 679 & -236 & 443 & 263 & -416 & -180 & 384 & -295 & -59 \\
49 & 20605 & 1.016 $\pm$ 0.110 & 644 & -312 & 332 & 585 & -58 & 254 & 707 & 63 & 375 \\
50 & 20745 & 1.041 $\pm$ 0.053 & 660 & -245 & 415 & 456 & -204 & 41 & 586 & -74 & 171 \\
51 & 20762 & 1.116 $\pm$ 0.061 & 707 & -236 & 472 & 381 & -326 & -90 & 508 & -199 & 36 \\
52 & 20827 & 1.137 $\pm$ 0.085 & 721 & -333 & 388 & -101 & -822 & -489 & 17 & -703 & -370 \\
53 & 21138 & 1.075 $\pm$ 0.161 & 681 & -234 & 447 & 261 & -420 & -187 & 385 & -297 & -63 \\
54 & 21256 & 1.090 $\pm$ 0.079 & 691 & -239 & 452 & 464 & -227 & 11 & 575 & -116 & 123 \\
55 & 21261 & 1.102 $\pm$ 0.060 & 699 & -234 & 465 & 273 & -426 & -192 & 388 & -311 & -77 \\
56 & 21723 & 1.129 $\pm$ 0.061 & 716 & -237 & 478 & 599 & -116 & 121 & 717 & 1 & 239 \\
57 & 22177 & 1.076 $\pm$ 0.071 & 682 & -237 & 444 & 190 & -492 & -254 & 310 & -372 & -134 \\
58 & 22253 & 1.123 $\pm$ 0.084 & 712 & -248 & 463 & 216 & -496 & -247 & 324 & -388 & -140 \\
59 & 22271 & 1.109 $\pm$ 0.051 & 703 & -224 & 478 & -200 & -903 & -678 & -94 & -796 & -572 \\
60 & 22654 & 1.129 $\pm$ 0.075 & 716 & -248 & 468 & 216 & -500 & -252 & 322 & -394 & -146 \\
61 & 23312 & 1.137 $\pm$ 0.106 & 721 & -309 & 412 & 110 & -611 & -302 & 226 & -495 & -186 \\
62 & 13806 & 1.132 $\pm$ 0.037 & 717 & -318 & 399 & -484 & -1201 & -883 & -362 & -1079 & -761 \\
63 & 13976 & 1.137 $\pm$ 0.060 & 721 & -292 & 429 & 323 & -397 & -106 & 467 & -253 & 38 \\
64 & 19786 & 1.080 $\pm$ 0.071 & 684 & -514 & 171 & -162 & -846 & -333 & -31 & -715 & -201 \\
\bottomrule
 \end{tabular} \\}
 \small
 \begin{justify}
\textbf{Notes.}
Columns are HIP: Hipparcos number; $M/R$: stellar mass over radius in solar units; $GR$: estimated gravitational redshift; $C$: estimated convective shift; $GR+C$: gravitational plus convective shift;
 $\Delta RV_{\rm M02}$: RV difference between HARPS (without zero point correction) and M02 measurements; $\Delta RV_{\rm M02}^{GR}$: $\Delta RV_{\rm M02}$ corrected from $GR$; $\Delta RV_{\rm M02}^{GR+C}$: $\Delta RV_{\rm M02}$ corrected from $GR+C$. $\Delta RV_{\rm G17}$: RV difference between HARPS (without zero point correction) and G17 measurements; $\Delta RV_{\rm G17}^{GR}$: $\Delta RV_{\rm G17}$ corrected from $GR$; $\Delta RV_{\rm G17}^{GR+C}$: $\Delta RV_{\rm G17}$ corrected from $GR+C$.
 \end{justify}
\end{table*}

\setcounter{table}{1}
\begin{table*}
 \caption{Continued.} 
 \centering
 \begin{tabular}{ccc|ccc|ccc|ccc}
\toprule
\# & HIP & $M/R$ & $GR$ & $C$ & $GR+C$ & $\Delta RV_{\rm M02}$ & $\Delta RV_{\rm M02}^{GR}$ & $\Delta RV_{\rm M02}^{GR+C}$ & $\Delta RV_{\rm G17}$ & $\Delta RV_{\rm G17}^{GR}$ & $\Delta RV_{\rm G17}^{GR+C}$ \\
 & & & (m s$^{-1}$) & (m s$^{-1}$) & (m s$^{-1}$) & (m s$^{-1}$) & (m s$^{-1}$) & (m s$^{-1}$) & (m s$^{-1}$) & (m s$^{-1}$) & (m s$^{-1}$) \\
\midrule
65 & 19793 & 1.086 $\pm$ 0.057 & 688 & -520 & 168 & 321 & -367 & 153 & 444 & -244 & 276 \\
66 & 19934 & 1.126 $\pm$ 0.042 & 714 & -353 & 361 & 419 & -294 & 58 & 544 & -170 & 183 \\
67 & 20082 & 1.137 $\pm$ 0.067 & 721 & -283 & 438 & 249 & -472 & -190 & 374 & -347 & -64 \\
68 & 20130 & 1.112 $\pm$ 0.052 & 705 & -384 & 321 & 378 & -327 & 57 & 498 & -207 & 177 \\
69 & 20237 & 1.047 $\pm$ 0.045 & 663 & -596 & 68 & -164 & -827 & -231 & -41 & -704 & -109 \\
70 & 20485 & 1.092 $\pm$ 0.083 & 692 & -235 & 457 & -771 & -1463 & -1228 & -651 & -1343 & -1109 \\
71 & 20563 & 1.132 $\pm$ 0.076 & 718 & -256 & 461 & 330 & -388 & -131 & 449 & -268 & -12 \\
72 & 20577 & 1.063 $\pm$ 0.252 & 674 & -581 & 93 & -1513 & -2187 & -1606 & -1388 & -2062 & -1481 \\
73 & 20741 & 1.088 $\pm$ 0.066 & 690 & -514 & 175 & -1161 & -1850 & -1336 & -1039 & -1728 & -1214 \\
74 & 20815 & 1.038 $\pm$ 0.039 & 658 & -615 & 43 & -383 & -1041 & -426 & -256 & -914 & -299 \\
75 & 20899 & 1.068 $\pm$ 0.058 & 677 & -567 & 109 & -445 & -1122 & -555 & -323 & -1000 & -432 \\
76 & 20951 & 1.129 $\pm$ 0.079 & 715 & -318 & 397 & 292 & -423 & -105 & 416 & -300 & 19 \\
77 & 21317 & 1.076 $\pm$ 0.066 & 682 & -523 & 159 & 59 & -623 & -100 & 177 & -505 & 18 \\
\bottomrule
 \end{tabular}
\end{table*}

\subsection{Cluster expansion}\label{sec:clexpantion}

\citet{Dravins99} pointed out that the astrometric RV are computed neglecting the hypothesis that the cluster may 
be expanding in the present, and under the assumption of constant expansion with time,
this would provide for the Hyades an upper limit of 70~m~s$^{-1}$. This implies that if cluster expansion did occur, the measured
 astrometric RVs should be systematically larger than the ones without expansion by up to this amount. This implies that 
 $\Delta RV_{\rm ZP}$ could be upper limits as far as this effect is concerned.

\subsection{Stellar activity}\label{sec:activity}

 As mentioned in Sect.~\ref{sec:m1}, enhanced stellar activity may induce a systematic effect, that is not considered in the zero point corrections. 
\citet{Lanza16} observed that the HARPS RV of the Sun redshifts by a few m~s$^{-1}$
when activity increases with the 11 years solar cycle. 
This effect, if extrapolated to active stars as the Hyades, would produce a systematic redshift of the stellar lines.
Note that this is not the RV Jitter induced by chromospheric variable activity on, e.g., a rotational period, 
that we assume will average out by observing many stars.
The spectra of all the stars would be systematically shifted because, assuming that 
the active stars are dominated by structures similar to those producing the 11 yrs solar cycle, 
the photospheric lines are systematically redshifted with respect to the Sun (our zero point reference), 
that is a relatively quiet and old star. 

All the RV studies we are aware of show a correlation between activity and spectroscopic 
redshift, although a direct link between activity indicators and 
shifts is not always present \citep{Dumusque12,Lovis11,Lanza16}.
The chromospheric activity level of the Hyades solar stars is about 4--5 times higher than the Sun \citep{Pace04}.
If we consider that the Sun varied by $\sim$4~m~s$^{-1}$ peak to peak over the last cycle (with a corresponding chromospheric activity variation of 
 $\sim 40 \%$ as measured from the CaII Index, \citealt{Lanza16}), and assuming a linear relationship between RV redshift and 
 the Logarithm of the Calcium H and K chromospheric flux, as found by \citet{Dumusque12} for $\alpha$~Cen~B, 
 we would expect that Hyades spectral RVs are affected by a systematic 
 zero point redshift of less than $\sim$ 50~m~s$^{-1}$.
 At present we cannot establish a precise value for this shift, also because the magnetic structures 
 (chromospheric network, plages, spots) contribute differently to the shift \citep{Haywood16}, 
 and we do not know in detail which structures are present on the surface of these stars. 

\subsection{Stellar rotation}\label{sec:rotation}

\begin{figure}
\centering
\includegraphics[width=0.45\textwidth]{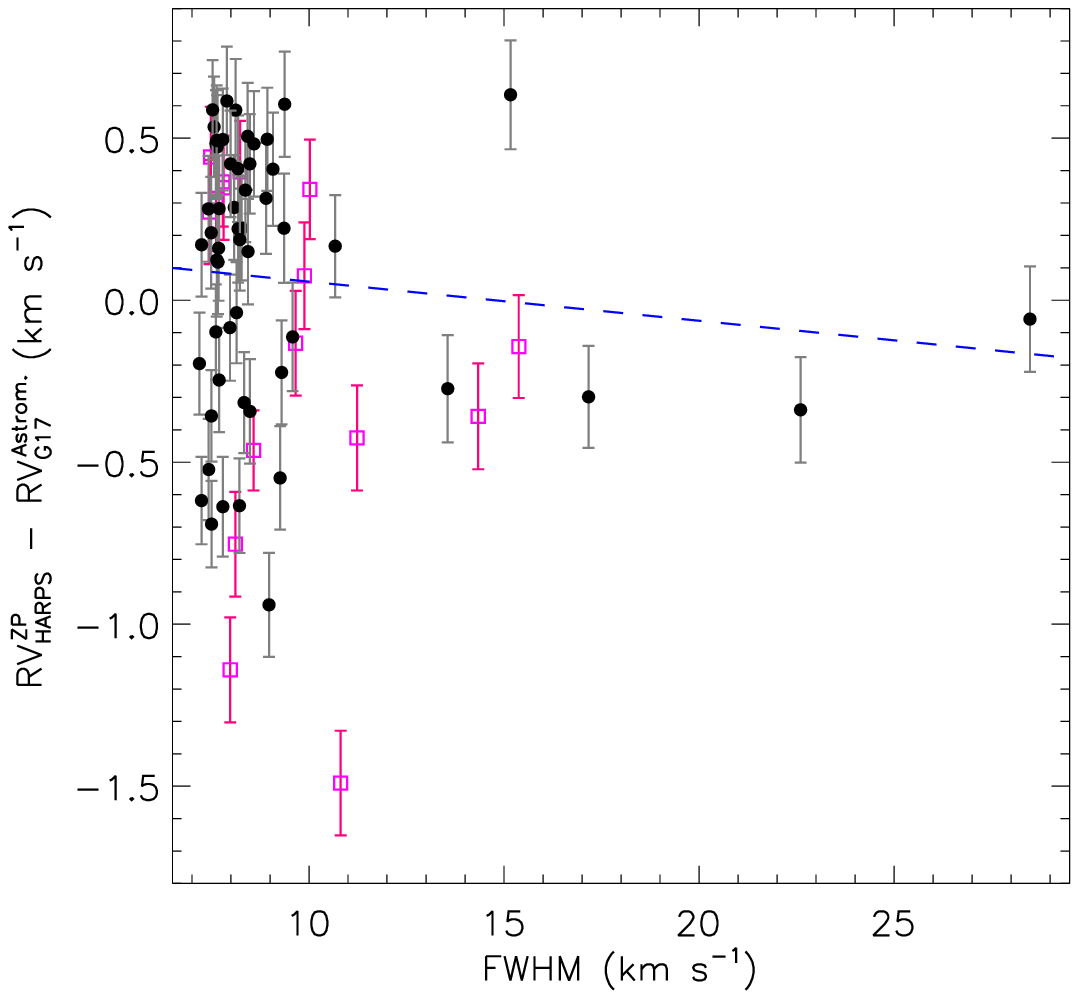}
 \caption{Difference between the spectroscopic and astrometric RV as a function of FWHM for our stars.
 Symbols and colours follow the same definition as in Fig.~\ref{fig:bvvsdelrv}. Dashed blue line is a linear fit with the data (excluded the BY~Dra variables).}
\label{fig:fwhmvsdelrv}
\end{figure}

One motivation for this study was the hint in M02 that the difference between astrometric and 
spectroscopic RVs might depend on stellar rotation.
\citet{Lindegren03} discuss the effects of enhanced rotation
on spectroscopic RV measurements and argue that some spectral lines
could shift as much as hundreds of m~s$^{-1}$.
We used as a proxy of the stellar rotational velocity the FWHM of the CCF, which is expected to be almost linearly related to the projected rotational velocity $v\sin i$ \citep[e.g.,][]{Geller10}.
Figure~\ref{fig:fwhmvsdelrv} shows this analysis within the FWHM span of our final 
sample, between 6 and 30~km~s$^{-1}$.
The Pearson correlation of this distribution (excluding BY~Dra variables) is low (-0.13) and
a linear fit with the data, showed by the dashed blue line, gives:
$$ \Delta RV_{\rm G17}^{\rm ZP} = - (0.012 \pm 0.006) \times {\rm FWHM} + (0.18 \pm 0.06) {\rm ~~ km~s}^{-1}, $$
which decreases slightly for increasing FWHM.
The relatively small slope (with 50\% uncertainty) may be caused by a selection effect because it is mainly produced by the fast rotators, which have a sparse distribution and relatively high spread. Indeed, the slope changes to $3 \pm 3$~m~s$^{-1}$ if we compute another linear fit for a restricted range of FWHM~$<$~12~km~s$^{-1}$, where the Pearson correlation is almost null (0.02).
We do not find therefore any measurable effect of rotational velocity on RV measurements in the range of stars studied, 
though some variations, dependent on the spectral type of the stars, is expected from our 3D models.

\subsection{Relativistic corrections}\label{sec:relcorr}

All the spectroscopic RV should be in principle corrected for a relativistic 
effect. One can estimate \citep[e.g.,][]{Lindegren03} that the difference between 
the spectroscopic and astrometric RV measurements is given by
$$ \left|RV_{\rm Spectr.} - RV_{\rm Astrom.}\right| \approx \frac{RV^2}{2 c}, $$
with $RV$ representing the ideal value free from any effect.
Considering an average RV of the cluster stars of 39.4~km~s$^{-1}$, 
this would imply a typical correction of $\sim$2.6~m~s$^{-1}$,
that can be considered negligible with the present level of accuracy. 
The relativistic correction for the transversal motion ($\sim$29~km~s$^{-1}$)
provides similar values.

\subsection{Galactic gravitational potential\label{sec:galpot}}

\begin{figure}
\centering
\includegraphics[width=\columnwidth]{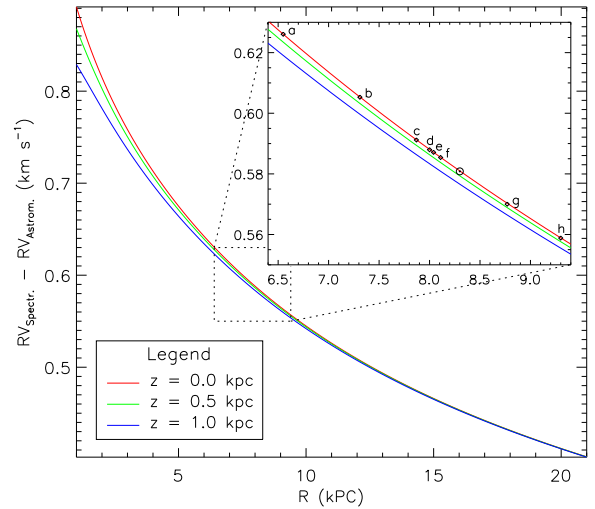}
 \caption{The expected effect of the Galactic potential on the difference between the spectroscopic 
 and astrometric RV measurements as a function of the cylindrical coordinates $R$ and $z$ with 
 the origin at the Galactic centre. A close view of the solar neighborhood is depicted in the 
 subpanel at the upper right region of the figure. A small list of clusters selected from \citet{Dravins99}
is shown in this subpanel by the letters ``a'' to ``h''. These are (a) NGC 6618, (b) IC 4725, (c) Upper Scorpius, 
(d) Coma Berenices, (e) Hyades, (f) Pleiades, (g) NGC 7789, and (h) NGC 2099. The Sun is illustrated by the solar symbol.}
\label{fig:galpot}
\end{figure}

The difference between astrometric and spectroscopic RV will also depend on the 
gravitational potential of the Galaxy, or better on the difference between the Galactic potential 
at the position of the observed star and of the sun.
 In principle, if our measurement were accurate enough, the comparison between astrometric and spectroscopic RV should trace
the Galactic potential. 
This difference is given by just GR, so it will depend on (M/R), where here M and R refer to the Mass and Radius of the Galaxy 
at the stellar position and at the solar position; such an effect will be very small for nearby stars, but it may be of 
several hundreds of m~s$^{-1}$ close to the bulge on in the MCs as pointed out by \citet{Lindegren03}. 

Gravitational redshift due to the Galactic potential
will only affect the spectroscopic measurements, not the astrometric ones. 
In other words, the same cluster, observed at 7~kpc distance towards the centre and the anti-centre, 
would show difference between spectroscopic and astrometric RVs of about 0.5~km~s$^{-1}$.
This can be seen clearly in Fig.~\ref{fig:galpot}, in which we used a model of the Galactic potential \citep{Piffl14}
to compute the expected effect. 
In the same figure we position a number of open clusters for which astrometric RV could be in principle obtained by 
a precise astrometric mission, such as {\em Gaia}. The maximum difference 
is of about 60~m~s$^{-1}$, which would be easily measurable as far as spectroscopic velocities are concerned, in particular if stars of similar spectral type were 
observed in the two clusters. We note, however, that to estimate accurately the valocity vector is not trivial, even for
 a nearby cluster as the Hyades (the reported G17 uncertainty is of 120~m~s$^{-1}$) and, in addition, that for the further clusters the astrometric precision 
will be lower and may not be sufficient even with {\em Gaia}. For some of the clusters, cluster expansion effect may also become relevant. 
\citet{Dravins99} compiled a list of open clusters that could be characterized by a {\em Gaia}-type mission and only for three the velocity vector should be meaurable to better than 100~m~s$^{-1}$. 
 As far as the Hyades are concerned, with a distance from the sun of 46.5~parsec \citep{vanLeeuwen07} and 
 a radial distance of ${8\,042.6}$~parsec the expected redshift produced by the galactic potential 
 is of 587 m~s$^{-1}$, whereas for the Sun it is of 581~m~s$^{-1}$,
 which gives a systematic difference of only 6 m~s$^{-1}$.

\section{Gravitational redshift and convection}\label{sec:maineffects}

\begin{figure}
\centering
\includegraphics[width=0.45\textwidth]{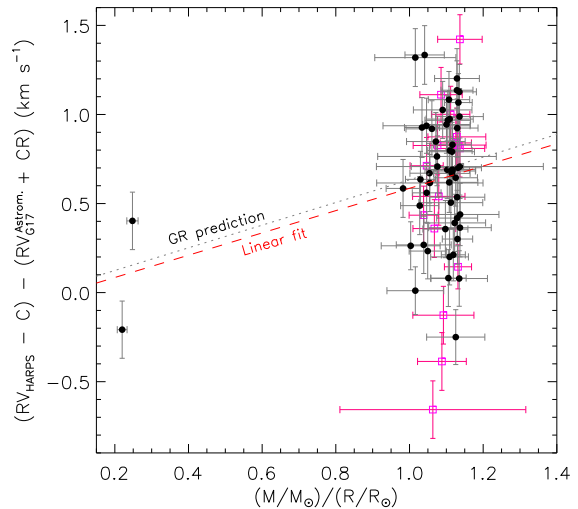}
\caption{Difference between the spectroscopic (corrected for convective shifts) and astrometric (corrected for cluster rotation)
 RV as a function of $M/R$ for the 71 stars of our sample.
 Target symbols and colours follow the same description as in Fig.~\ref{fig:bvvsdelrv}. 
 The grey dotted line depicts the theoretical prediction of the GR from Eq.~(\ref{eq:gr}).
 The red dashed line is a linear fit to the observed data, excluding the BY~Dra variables.}
\label{fig:mrvsdelrv}
\end{figure}

As introduced in the previous sections, GR and convection (described in Sects.~\ref{sec:gravrsh} and~\ref{sec:convshift} respectively)
are the two effects that dominate the distortions in spectroscopic RV measurements.
Even if the two effects are comparable and have opposite sign, we do not expect that they cancel completely, as can be seen from the theoretical values
given in Table~\ref{table:shifts}.

As shown above, the Hyades cluster rotates, and this effect has not been taken into account in the astrometric measurements. In order to be consistent and
to perform a detailed analysis, we have added to the astrometric velocity of our 71 stars the contribution of cluster rotation, using the gradient found in Sect.~\ref{sec:asymmetry}. Hipparcos parallaxes were used for the 6 targets with missing G17 parallaxes.

If we correct $\Delta RV_{\rm G17}$ (the simple difference between measured HARPS and G17 astrometric velocities)
for cluster rotation (CR) and for the convective shift,
the resulting RV difference, namely $\Delta RV_{\rm G17}^{C+CR}$, is expected to be dominated by GR,
which can thus be determined empirically.
Figure~\ref{fig:mrvsdelrv} shows $M/R$ versus $\Delta RV_{\rm G17}^{C+CR}$ for our 71 stars. 
The two Hyades giants show smaller $\Delta RV_{\rm G17}^{C+CR}$ than the average of the other stars, as expected by GR
because of their substantially smaller $M/R$.
A linear fit of $M/R$ versus $\Delta RV_{\rm G17}^{C+CR}$ data, excluded the BY~Dra variables, provides:
$$ \Delta RV_{\rm G17}^{C+CR} = (0.626 \pm 0.131) \times (M/R) - (0.040 \pm 0.140) {\rm ~~ km~s}^{-1}. $$
The slope of $\Delta RV_{\rm G17}^{C+CR}$ of $626 \pm 131$~m~s$^{-1}$ agrees well with the theoretical prediction of GR (grey dotted line in Fig.~\ref{fig:mrvsdelrv}),
given by Eq.~(\ref{eq:gr}), within the uncertainty of the measurement.
The 131~m~s$^{-1}$ uncertainty is somewhat high because only two giants are part of our sample, and they 
are essential to extend the $M/R$ versus $\Delta RV_{\rm G17}^{C+CR}$ distribution to low $M/R$ values for computing and determine the GR slope.

\begin{figure}
\centering
\includegraphics[width=0.45\textwidth]{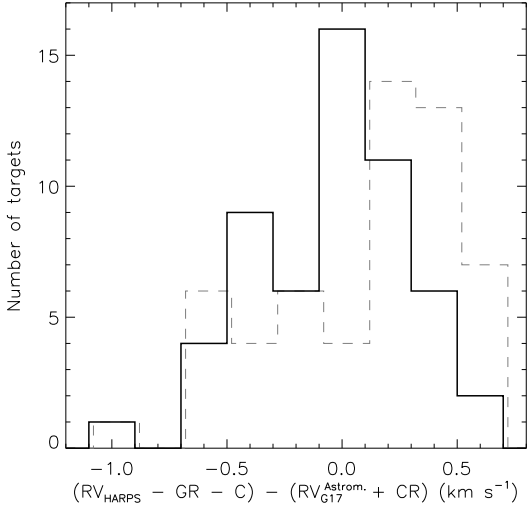}
\caption{Solid line is the distribution of the difference between
 spectroscopic (corrected for convective shifts and gravitational redshift) and astrometric (corrected for cluster rotation) RV for 55 stars of our sample
 (the BY Dra have been excluded). The distribution is gaussian and centered very close to zero.
 Dashed line is the distribution of the RV difference for the same sample without correction for comparison (with a slight shift of +0.02~km~s$^{-1}$ for better visualization), where there is a clear asymmetry as in Fig.~\ref{fig:histdelrv}.}
\label{fig:final}
\end{figure}

We finally show the results of our best estimate for the comparison between astrometric and spectroscopic RVs in Fig.~\ref{fig:final}.
The distribution represents, for the stars of our sample (excluding the BY Dra), the difference between the spectroscopic RVs (subtracted by the GR and convective contribution) and the astrometric G17 RVs (with the cluster rotation added). As anticipated in Sect.~\ref{sec:asymmetry}, introducing
the cluster rotation brings the distribution close to a normal one. The mean difference is of $-33$~m~s$^{-1}$, the median $-16$~m~s$^{-1}$ with a $\sigma$ of 
$347$~m~s$^{-1}$. This result shows quite a good agreement between spectroscopic and astrometric radial velocities,
indirectly validating the steps and models used.

\begin{table*}
\caption{Summary of $\Delta RV$ results.}
\label{tab:summary}
{\centering
\begin{tabular}{c|cc|cccc|cccc}
\toprule
\multicolumn{11}{c}{Major effects} \\
\midrule
 & $GR$ & $C$ & $\Delta RV_{\rm M02}$ & $\Delta RV_{\rm M02}^{\rm ZP}$ & $\Delta RV_{\rm M02}^{GR}$ & $\Delta RV_{\rm M02}^{GR+C}$ & $\Delta RV_{\rm G17}$ & $\Delta RV_{\rm G17}^{\rm ZP}$ & $\Delta RV_{\rm G17}^{GR}$ & $\Delta RV_{\rm G17}^{GR+C}$ \\
 & (m s$^{-1}$) & (m s$^{-1}$) & (m s$^{-1}$) & (m s$^{-1}$) & (m s$^{-1}$) & (m s$^{-1}$) & (m s$^{-1}$) & (m s$^{-1}$) & (m s$^{-1}$) & (m s$^{-1}$) \\
\midrule
 Mean & $678$ & $-366$ & $+17$ & $-85$ & $-661$ & $-295$ & $+139$ & $+37$ & $-538$ & $-172$ \\
 Median & $+701$ & $-327$ & $+150$ & $+48$ & $-530$ & $-189$ & $+273$ & $+171$ & $-404$ & $-64$ \\
 $\sigma$ & $95$ & $130$ & $464$ & $464$ & $443$ & $429$ & $462$ & $462$ & $441$ & $426$ \\
\bottomrule
\end{tabular} \\
\begin{tabular}{cccc}
\toprule
\multicolumn{4}{c}{Minor effects} \\
\midrule
 Cluster & Stellar & General & Galactic \\
 Expansion & Activity & Relativity & Potential \\
 (m s$^{-1}$) & (m s$^{-1}$) & (m s$^{-1}$) & (m s$^{-1}$) \\
\midrule
 $> -70$ & $< +50$ & $< +3$ spectra & $+6$ \\
 & & $< +3$ astro & \\
\bottomrule
\end{tabular} \\}
 \small
 \begin{justify}
\textbf{Notes.}
The top panel shows the major effects affecting
the RV measurements. Their mean and median values, and standard deviations are
obtained from the data of Table~\ref{table:shifts}. Columns named $GR$ and $C$
refers to the gravitational redshift and convective shift,
as discussed in Sects.~\ref{sec:gravrsh} and~\ref{sec:convshift}, respectively.
$\Delta RV_{\rm M02}$ and $\Delta RV_{\rm M02}^{\rm ZP}$ refer to the HARPS minus M02
astrometric RV difference without and with zero point correction, respectively. 
Column $\Delta RV_{\rm M02}^{GR}$ shows the global
shift on the $\Delta RV_{\rm M02}$ values when the GR contribution is subtracted,
whereas $\Delta RV_{\rm M02}^{GR+C}$ gives $\Delta RV_{\rm M02}$
corrected for the two most relevant effects: GR and convective shift $C$.
Columns from $\Delta RV_{\rm G17}$ to $\Delta RV_{\rm G17}^{GR+C}$ follow the same descriptions as those for columns $\Delta RV_{\rm M02}$ to $\Delta RV_{\rm M02}^{GR+C}$, but using the G17 astrometric RVs.
The bottom panel shows the minor effects,
 discussed in Sects.~\ref{sec:clexpantion}, \ref{sec:activity}, and \ref{sec:relcorr}, respectively.
A positive value means that the effect should be subtracted from $\Delta RV$, because it
contributes positively to the difference, so either it adds
a redshift to the spectroscopic measurement, or decreases the astrometric one.
 \end{justify}
\end{table*}

\section{Hyades RV} \label{sec:averagervs}

The latest Hyades RV in literature before G17 is $39.4 \pm 0.6$~km~s$^{-1}$ \citep{Maderak13}, computed by averaging the RV of their sample.
These authors also measured a median difference of $-0.56$~km~s$^{-1}$ between their RV measurements and those published by P98, similar to our findings. 
G17 provide a new estimate of the Hyades RV of $39.38 \pm 0.16$ km~s$^{-1}$ (purely astrometric solution) computed from kinematic models.

The new HARPS measurements, coupled with our accurate zero points and the discovery of more double or RV variable stars, 
can be used to provide a new, accurate spectroscopic RV for this cluster, namely $RV_{\rm Hyades}^{\rm HARPS}$.
To do this, we considered the 3D structure of the cluster by using the equatorial coordinates of the stars and G17 parallaxes.
Then, we analyzed a few subsample selections of stars within a 3--10~pc radius from the cluster centre, 
namely between the cluster core and its tidal radius \citep{Madsen03}.
We used the Hyades central coordinates provided in G17.
The two giants of our sample were excluded because the GR effect makes them to deviate from the dwarfs and then to bias our results.
For each subsample selection, the stellar RVs corrected for zero-point were adjusted
to their projection to the cluster centre, by $RV/\cos\theta$, where $\theta$ is the angular 
distance to the cluster centre.
Since the projected RV distribution is not fully gaussian, but slightly asymmetric,
we use the median value of the projected RVs as a suitable measure of $RV_{\rm Hyades}^{\rm HARPS}$.

The median value of the projected HARPS RVs tends to increase the smaller the subsample region is.
The selection within the tidal radius ($<$10~pc, comprising 46 stars) provides $RV_{\rm Hyades}^{\rm HARPS} = 39.07 \pm 0.19$~km~s$^{-1}$, whereas the
cluster core ($<$3~pc, that includes 17 targets) provides $RV_{\rm Hyades}^{\rm HARPS} = 39.36 \pm 0.26$~km~s$^{-1}$. 
We assume the latter value best represents the Hyades RV as being its centroid RV.

\section{Conclusions} \label{sec:conclusion}

We summarise in Table~\ref{tab:summary} the results of our comparative analysis of the difference between the astrometric 
and spectroscopic RVs of the Hyades stars. 
After cleaning the sample from RV variable of different types our sample consists of 71 late type stars. 
The distribution of RV differences (HARPS -- astrometric, zero-point corrected), $\Delta RV_{\rm M02}^{\rm ZP}$ and $\Delta RV_{\rm G17}^{\rm ZP}$, are not gaussian, 
but has a negative tail. The first important result is that, independent of the solution and zero point used, the 
difference is rather small: $-85$ or $+48$ m~s$^{-1}$ 
for $\Delta RV_{\rm M02}^{\rm ZP}$ and $+37$ or $+171$ for $\Delta RV_{\rm G17}^{\rm ZP}$, 
depending whether the mean or the median are assumed, respectively.

 The agreement between the two methods (Doppler vs. astrometric) is still remarkable, when considering that many effects are expected to act on the spectroscopic RVs, and modify them of up to many hundreds m~s$^{-1}$.
Our analysis also allowed us to determine the rotation of the Hyades cluster, at 42.3~m~s$^{-1}$pc$^{-1}$. This rotation is the same given in
 \citet{Vereshchagin13a}, but our new measurement has higher statistical significance.

We consider the main factors affecting the measurements accuracy and we model the two major 
ones, stellar gravitational redshift and convective motions.
After modeling the two effects, we retrieve a zero point correction for the HARPS 
G2 mask that is in agreement with the empirical one by better than 50~m~s$^{-1}$. After applying the 
 two corrections to the whole sample, and the correction for cluster rotation, the skewness in the $\Delta$RV distribution disappears,
 and the agreement between spectroscopic and astrometric radial velocities becomes very good: $-16$~m~s$^{-1}$ (median), with a $\sigma$ of 347 m~s$^{-1}$, that is very close to the velocity dispersion of the
 cluster as measured from proper motions.
This result shows that it is possible to determine accurate RVs.
 
The giants have more blueshifted spectra than the dwarfs, as expected from gravitational redshift, 
and we could verify the scaling of GR with stellar mass and radius for the first time. Gravitational redishift has been previously observed in white dwarfs and measured on the Sun \citep[e.g.,][]{hol10,for61},
 but never in an open cluster.
After cleaning the HARPS RVs for the convection effects and correcting the astrometric RVs for cluster rotation, the resulting difference with the G17 RVs,
namely $\Delta RV_{\rm G17}^{C+CR}$, is expected to be only dominated by GR,
which can thus be measured from the $M/R$ versus $\Delta RV_{\rm G17}^{C+CR}$ relation. The slope of $626 \pm 131$~m~s$^{-1}$
for this relation is fully compatible with the one expected theoretically (of 636.5~m~s$^{-1}$) from Eq.~(\ref{eq:gr}).

All other effects considered could affect this cluster up to a maximum of 70~m~s$^{-1}$. Enhanced chromospheric 
activity would systematically redshift the spectra of less than 50~m~s$^{-1}$, while a potential cluster expansion would redshift the 
astrometric ones systematically by up to 70~m~s$^{-1}$.
According to our models, stellar rotational velocity influences the spectroscopic RV measurement (with shifts that depend on the 
spectral type), but we cannot detect clear systematic differences when comparing $\Delta RV_{\rm G17}^{\rm ZP}$ against this paramater.
It is finally interesting to note that all these effects provide a contribution much smaller than 500~m~s$^{-1}$, 
which is the difference that could be expected, for instance by comparing astrometric and spectroscopic radial velocities in
clusters located in the inner region of the 
Galaxy with clusters located in the outer one.
Thus the comparison between spectroscopic and astrometric RV could trace the galactic potential in these areas.

\section*{Acknowledgements}

We thank the referee, Prof. S. Shectman, for very useful comments to the original manuscript. 
ICL acknowledges a Post-Doctoral fellowship from the CNPq Brazilian agency (Science Without Borders program, Grant No. 207393/2014-1).
LPA acknowledges support from brazilian PVE programme.
HGL acknowledges financial support by the Sonderforschungsbereich SFB\,881
``The Milky Way System'' (subproject A4) of the German Research Foundation
(DFG).
Research activities of the Observational Astronomy Board of the Federal University of Rio Grande do Norte (UFRN) are supported by continuous grants from the CNPq and FAPERN Brazilian agencies.





\begin{thebibliography}{99}

\bibitem[Allende Prieto et al.(2009)]{2009MmSAI..80..622A} Allende Prieto, 
C., Koesterke, L., Ram{\'{\i}}rez, I., Ludwig, H.-G., 
\& Asplund, M.\ 2009, Memorie della Societa Astronomica Italiana, 80, 622 

\bibitem[Beckers(1977)]{Beckers77} Beckers, J. M. 1977, ApJ, 213, 900

\bibitem[B\"ohm-Vitense et al.(2002)]{BohmVitense02} B\"ohm-Vitense, E., Robinson, R., Carpenter, K. et al. 2002, ApJ, 569, 941

\bibitem[Bressan et al.(2012)]{Bressan12} Bressan, A., Marigo, P., Girardi, L. et al. 2012, MNRAS, 427, 127

\bibitem[Chen et al.(2015)]{Chen15} Chen, Y., Bressan, A., Girardi, L. 2015, MNRAS, 452, 1068

\bibitem[Chen et al.(2014)]{Chen14} Chen, Y., Girardi, L., Bressan, A. et al. 2014, MNRAS, 444, 2525


\bibitem[Cochran et al.(2002)]{Cochran02} Cochran, W. D., Hatzes, A. P., \& Paulson, D. B. 2002, AJ, 124, 565

\bibitem[de Bruijne(2001)]{deBruijne01} de Bruijne, J. H. J., Hoogerwerf, R., \& de Zeeuw, P. T. 2001, A\&A, 367, 111

\bibitem[de Bruijne et al.(2002)]{deBruijne02} de Bruijne, J. H. J., Reynolds, A. P., Perryman, M. A. C. et al. 2002, A\&A, 381, L57

\bibitem[Dumusque et al.(2012)]{Dumusque12} Dumusque, X., Pepe, F., Lovis, C. et al. 2012, Nature, 491, 207

\bibitem[Dutra-Ferreira et al.(2016)]{DutraFerreira16} Dutra-Ferreira, L., Pasquini, L., Smiljanic, R. et al. 2016, A\&A, 585, A75

\bibitem[Dravins et al.(1999)]{Dravins99} Dravins, D., Lindegren, L., \& Madsen, S. 1999, A\&A, 348, 1040

\bibitem[Einstein(1917)]{Einstein17} Einstein, A. 1917, SPAW, 142

\bibitem[Forbes(1961)]{for61} Forbes, E.G. 1961, {\em The problem of the solar red shifts}, University of St. Andrews (United Kingdom).

\bibitem[Gaia Collaboration et al.(2017)]{gaia17} Gaia Collaboration, van Leeuwen, F., Vallenari, A. et al. 2017, A\&A, 601, A19 [G17]

\bibitem[Geller et al.(2010)]{Geller10} Geller, A. M., Mathieu, R. D., Braden, E. K. et al. 2010, AJ, 139, 1383

\bibitem[Grevesse \& Sauval(1998)]{Grevesse98} Grevesse, N., \& Sauval, A. J. 1998, SSRv, 85, 161

\bibitem[Griffin(2012)]{Griffin12} Griffin, R. F. 2012, JApA, 33, 29

\bibitem[Griffin et al.(1988)]{Griffin98} Griffin, R. F., Griffin, R. E. M., Gunn, J. E. et al. 1988, AJ, 96, 172

\bibitem[Guenther et al.(2005)]{Guenther05} Guenther, E. W., Paulson, D. B., Cochran, W. D. et al. 2005, A\&A, 442, 1031

\bibitem[Hartkopf et al.(2001)]{Hartkopf01} Hartkopf, W. I., McAlister, H. A., \& Mason, B. D. 2001, AJ, 122, 3480

\bibitem[Haywood et al.(2016)]{Haywood16} Haywood, R. D., Collier Cameron, A., Unruh, Y. C. et al. 2016, MNRAS, 457, 3637


\bibitem[Holberg(2010)]{hol10} Holberg, J. B. 2010, JHA, 41, 41

\bibitem[Kurucz \emph{et al.}(1984)]{Kurucz84}Kurucz, R.L.,
 Furenlid, I., Brault, J., and Testerman, L.: 1984, {\it National Solar
 Observatory Atlas, Sunspot, New Mexico: National Solar Observatory, 1984}

\bibitem[Lanza et al.(2016)]{Lanza16} Lanza, A. F., Molaro, P., Monaco, L., \& Haywood, R. D. 2016, A\&A, 587, A103

\bibitem[Lindegren \& Dravins(2003)]{Lindegren03} Lindegren, L., \& Dravins, D. 2003 A\&A, 401, 1185

\bibitem[Lindegren et al.(2000)]{Lindegren00} Lindegren, L., Madsen, S., \& Dravins, D. 2000, A\&A, 356, 1119

\bibitem[Lo Curto et al.(2015)]{LoCurto15} Lo Curto, G., Pepe, F., Avila, G. et al. 2015, Msngr, 162, 9

\bibitem[Lovis et al.(2011)]{Lovis11} Lovis, C., S\'egransan, D., Mayor, M. et al. 2011, A\&A, 528, A112

\bibitem[\protect\citeauthoryear{Ludwig, Freytag, \&
 Steffen}{1999}]{1999A&A...346..111L} Ludwig H.-G., Freytag B., Steffen M.,
 1999, A\&A, 346, 111

\bibitem[\protect\citeauthoryear{Ludwig et al.}{2017}]{Ludwig17} Ludwig,
 H.-G., Koesterke, L., Allende Prieto, C., Bertran de Lis, S., Freytag, B.,
 Caffau, E., 2017, in prep.

\bibitem[Maderak et al.(2013)]{Maderak13} Maderak, R. M., Deliyannis, C. P., King, J. R. et al. 2013, AJ, 146, 143

\bibitem[Madsen(2003)]{Madsen03} Madsen, S. 2003, A\&A 401, 565

\bibitem[Madsen et al.(2002)]{Madsen02} Madsen, S., Dravins, D., \& Lindegren, L. 2002, A\&A, 381, 446 [M02]

\bibitem[Mason et al.(2009)]{Mason09} Mason, B. D., Hartkopf, W. I., Gies, D. R. et al. 2009, AJ, 137, 3358

\bibitem[Mayor et al.(2003)]{Mayor03} Mayor, M., Pepe, F., Queloz, D. et al. 2003, Msngr, 114, 20

\bibitem[Molaro et al.(2016)]{Molaro16} Molaro, P., Lanza, A. F., Monaco, L. et al. 2016, MNRAS, 458, L54


\bibitem[Pace \& Pasquini(2004)]{Pace04} Pace, G., \& Pasquini, L. 2004, A\&A, 426, 1021

\bibitem[Pasquini et al.(2011)]{Pasquini11} Pasquini, L., Melo, C., Chavero, C. et al. 2011, A\&A, 526, A127



\bibitem[Paulson et al.(2004)]{Paulson04} Paulson, D. B., Saar, S. H., Cochran, W. D. et al. 2004, AJ, 127, 1644

\bibitem[Paulson et al.(2003)]{Paulson03} Paulson, D. B., Sneden, C., \& Cochran, W. D. 2003, AJ, 125, 3185

\bibitem[\protect\citeauthoryear{Pepe et al.}{2002}]{2002A&A...388..632P} Pepe
 F., Mayor M., Galland F., Naef D., Queloz D., Santos N.~C., Udry S., Burnet
 M., 2002, A\&A, 388, 632 

\bibitem[Perryman et al.(1998)]{Perryman98} Perryman, M. A. C., Brown, A. G. A., Lebreton, Y. et al. 1998, A\&A, 331, 81 [P98]

\bibitem[Piffl(2014)]{Piffl14} Piffl, T., Binney, J., McMillan, P. J. et al. 2014, MNRAS, 445, 3133

\bibitem[Poretti(2001)]{Poretti01} Poretti, E. 2001, A\&A, 371, 986

\bibitem[Poretti et al.(2009)]{Poretti09} Poretti, E., Michel, E., Garrido, R. et al. 2009, A\&A, 506, 85

\bibitem[Quinn et al.(2014)]{Quinn14} Quinn, S. N., White, R. J., Latham, D. W. et al. 2014, ApJ, 787, 27

\bibitem[Randich et al.(2000)]{Randich00} Randich, S., Pasquini, L., \& Pallavicini, R. 2000, A\&A, 356, L25

\bibitem[Reino et al.(2018)]{Reino18} Reino, S., de Bruijne, J., Zari, E. et al. 2018, MNRAS, 477, 3197

\bibitem[Saar \& Donahue(1997)]{Saar97} Saar, S. H., \& Donahue, R. A. 1997, ApJ, 485, 319


\bibitem[Tang et al.(2014)]{Tang14} Tang, J., Bressan, A., Rosenfield, P. 2014, MNRAS, 445, 4287

\bibitem[\protect\citeauthoryear{Tonry \& Davis}{1979}]{1979AJ.....84.1511T}
 Tonry J., Davis M., 1979, AJ, 84, 1511

\bibitem[van Leeuwen(2007)]{vanLeeuwen07} van Leeuwen, F. 2007, A\&A, 474, 653

\bibitem[van Leeuwen(2009)]{vanLeeuwen09} van Leeuwen, F. 2009, A\&A, 497, 209

\bibitem[Vereshchagin \& Chupina(2013)]{Vereshchagin13a} Vereshchagin, S.V., \& Chupina, N. V. 2013, AN, 334, 892


\end{thebibliography}





\appendix 

\section{Theoretical convective line shifts}
\label{a:vconv}

\begin{table*}
\caption[]{Cross-correlation results: for 21 synthetic spectra of 3D solar
 metallicity models labelled by ``Spec.ID'' and their atmospheric parameters
 (\Teff\ (K), logarithmic surface gravity (cm\,s$^{-2}$)) the table lists the
 FWHM of the cross-correlation peak (first line), and its position (second
 line). A negative value of the peak position corresponds to a
 blueshift. $\xi$ is the broadening parameter applied in the convolution of
 the synthetic spectra.\label{t:vconv}}
\begin{tabular}{rrrrrrrrrrrrrrrrrr}
\hline\noalign{\smallskip}
\multicolumn{2}{c}{Spec.~ID} & \multicolumn{16}{c}{FWHM of cross-correlation peak [km s$^{-1}$]} \\
\Teff & \logg & \multicolumn{16}{c}{Convective line shift [m s$^{-1}$]} \\
\multicolumn{2}{c}{$\xi$ [km s$^{-1}$] $\rightarrow$} & 0.0 & 1.0 & 2.0 & 2.44 & 3.0 & 4.0 & 5.0 & 6.0 & 7.0 & 8.0 & 9.0 & 10.0 & 12.5 & 15.0 & 17.5 & 20.0\\
\hline\noalign{\smallskip}
\multicolumn{2}{c}{hyd0001} & 6.1 & 6.4 & 7.2 & 7.7 & 8.4 & 9.8 & 11.4 & 13.2 & 15.1 & 17.0 & 19.0 & 20.9 & 26.0 & 31.4 & 36.7 & 42.3\\
3813 & 4.00 & -329 & -332 & -338 & -340 & -341 & -346 & -354 & -364 & -374 & -380 & -381 & -377 & -353 & -315 & -268 & -215\\
\multicolumn{2}{c}{hyd0003} & 8.2 & 8.4 & 8.8 & 9.1 & 9.6 & 10.5 & 11.7 & 13.0 & 14.5 & 16.0 & 17.6 & 19.2 & 23.4 & 27.6 & 31.8 & 36.1\\
4018 & 1.50 & -531 & -523 & -497 & -482 & -462 & -430 & -407 & -393 & -385 & -380 & -376 & -372 & -355 & -326 & -286 & -236\\
\multicolumn{2}{c}{hyd0007} & 6.2 & 6.6 & 7.4 & 7.9 & 8.6 & 10.1 & 11.6 & 13.5 & 15.4 & 17.4 & 19.4 & 21.3 & 26.5 & 31.9 & 37.3 & 42.8\\
3964 & 4.50 & -300 & -304 & -311 & -313 & -315 & -320 & -328 & -338 & -347 & -352 & -353 & -349 & -323 & -286 & -241 & -190\\
\multicolumn{2}{c}{hyd0010} & 7.1 & 7.3 & 7.8 & 8.1 & 8.5 & 9.5 & 10.7 & 12.1 & 13.5 & 15.1 & 16.7 & 18.3 & 22.5 & 26.9 & 31.1 & 35.3\\
4477 & 2.50 & -511 & -513 & -514 & -512 & -509 & -504 & -501 & -502 & -505 & -508 & -509 & -509 & -497 & -472 & -436 & -393\\
\multicolumn{2}{c}{hyd0012} & 6.5 & 6.8 & 7.5 & 7.9 & 8.6 & 9.9 & 11.4 & 13.1 & 14.8 & 16.6 & 18.4 & 20.2 & 25.1 & 30.1 & 35.0 & 40.1\\
4509 & 4.50 & -301 & -305 & -312 & -314 & -317 & -323 & -332 & -343 & -354 & -364 & -370 & -373 & -365 & -343 & -312 & -275\\
\multicolumn{2}{c}{hyd0016} & 8.3 & 8.4 & 8.9 & 9.1 & 9.5 & 10.4 & 11.5 & 12.7 & 13.9 & 15.3 & 16.8 & 18.3 & 22.1 & 26.1 & 30.0 & 33.9\\
4968 & 2.50 & -664 & -657 & -635 & -622 & -604 & -572 & -546 & -527 & -513 & -502 & -493 & -484 & -461 & -434 & -404 & -373\\
\multicolumn{2}{c}{hyd0020} & 6.6 & 6.8 & 7.3 & 7.6 & 8.1 & 9.2 & 10.4 & 11.8 & 13.3 & 14.8 & 16.5 & 18.2 & 22.5 & 26.8 & 31.0 & 35.3\\
4923 & 3.50 & -444 & -452 & -463 & -466 & -469 & -473 & -476 & -482 & -488 & -494 & -499 & -501 & -493 & -472 & -441 & -405\\
\multicolumn{2}{c}{hyd0024} & 6.2 & 6.4 & 7.0 & 7.4 & 7.9 & 9.1 & 10.4 & 11.8 & 13.4 & 15.1 & 16.8 & 18.5 & 23.0 & 27.5 & 31.9 & 36.3\\
4954 & 4.00 & -329 & -338 & -352 & -358 & -363 & -372 & -380 & -390 & -400 & -409 & -415 & -419 & -415 & -395 & -365 & -329\\
\multicolumn{2}{c}{hyd0028} & 6.2 & 6.4 & 7.1 & 7.5 & 8.1 & 9.3 & 10.6 & 12.2 & 13.8 & 15.5 & 17.3 & 19.1 & 23.7 & 28.4 & 32.9 & 37.5\\
4982 & 4.50 & -288 & -294 & -304 & -308 & -312 & -319 & -327 & -337 & -347 & -356 & -363 & -367 & -364 & -345 & -316 & -281\\
\multicolumn{2}{c}{hyd0032} & 7.0 & 7.2 & 7.7 & 8.0 & 8.4 & 9.4 & 10.5 & 11.8 & 13.2 & 14.6 & 16.1 & 17.7 & 21.7 & 25.7 & 29.6 & 33.6\\
5432 & 3.50 & -622 & -621 & -615 & -610 & -603 & -592 & -584 & -579 & -577 & -576 & -574 & -570 & -554 & -529 & -499 & -467\\
\multicolumn{2}{c}{hyd0036} & 6.4 & 6.6 & 7.1 & 7.4 & 7.9 & 9.0 & 10.2 & 11.5 & 13.0 & 14.5 & 16.1 & 17.7 & 21.9 & 26.0 & 30.1 & 34.2\\
5475 & 4.00 & -481 & -487 & -498 & -501 & -503 & -506 & -508 & -512 & -518 & -522 & -526 & -526 & -517 & -495 & -465 & -431\\
\multicolumn{2}{c}{hyd0040} & 6.1 & 6.3 & 6.9 & 7.2 & 7.8 & 8.9 & 10.2 & 11.6 & 13.1 & 14.7 & 16.4 & 18.1 & 22.4 & 26.8 & 31.0 & 35.3\\
5488 & 4.50 & -326 & -335 & -350 & -355 & -361 & -369 & -377 & -385 & -394 & -402 & -408 & -411 & -406 & -386 & -357 & -324\\
\multicolumn{2}{c}{hyd0044} & 7.7 & 7.9 & 8.3 & 8.6 & 9.0 & 10.0 & 11.0 & 12.3 & 13.5 & 14.8 & 16.2 & 17.6 & 21.3 & 25.1 & 28.8 & 32.6\\
5884 & 3.50 & -787 & -780 & -761 & -749 & -735 & -709 & -687 & -669 & -654 & -641 & -628 & -614 & -575 & -534 & -493 & -456\\
\multicolumn{2}{c}{hyd0048} & 6.8 & 7.0 & 7.5 & 7.8 & 8.3 & 9.2 & 10.4 & 11.6 & 13.0 & 14.4 & 15.9 & 17.4 & 21.2 & 25.1 & 29.0 & 32.8\\
5928 & 4.00 & -637 & -638 & -636 & -634 & -630 & -623 & -618 & -615 & -613 & -611 & -607 & -601 & -575 & -539 & -499 & -460\\
\multicolumn{2}{c}{hyd0052} & 6.3 & 6.5 & 7.0 & 7.3 & 7.8 & 8.9 & 10.1 & 11.5 & 12.9 & 14.4 & 16.0 & 17.6 & 21.6 & 25.8 & 29.8 & 33.8\\
5865 & 4.50 & -475 & -484 & -498 & -503 & -508 & -515 & -520 & -526 & -532 & -536 & -539 & -538 & -523 & -493 & -456 & -416\\
\multicolumn{2}{c}{hyd0057} & 7.4 & 7.6 & 8.0 & 8.3 & 8.7 & 9.7 & 10.8 & 12.0 & 13.2 & 14.6 & 16.0 & 17.4 & 21.0 & 24.8 & 28.5 & 32.3\\
6229 & 4.00 & -806 & -801 & -785 & -775 & -762 & -738 & -718 & -701 & -687 & -674 & -659 & -644 & -599 & -549 & -499 & -452\\
\multicolumn{2}{c}{hyd0061} & 6.6 & 6.8 & 7.3 & 7.6 & 8.0 & 9.0 & 10.2 & 11.5 & 12.9 & 14.3 & 15.8 & 17.3 & 21.1 & 25.0 & 28.8 & 32.7\\
6233 & 4.50 & -608 & -612 & -616 & -616 & -615 & -611 & -609 & -608 & -608 & -606 & -602 & -595 & -565 & -522 & -475 & -427\\
\multicolumn{2}{c}{hyd0065} & 8.2 & 8.4 & 8.8 & 9.1 & 9.5 & 10.3 & 11.4 & 12.6 & 13.7 & 14.9 & 16.2 & 17.6 & 21.1 & 24.7 & 28.3 & 32.1\\
6484 & 4.00 & -975 & -964 & -935 & -918 & -896 & -857 & -823 & -793 & -767 & -743 & -720 & -697 & -636 & -575 & -517 & -462\\
\multicolumn{2}{c}{hyd0069} & 6.9 & 7.1 & 7.5 & 7.8 & 8.3 & 9.3 & 10.4 & 11.7 & 12.9 & 14.3 & 15.7 & 17.2 & 20.9 & 24.6 & 28.4 & 32.2\\
6456 & 4.50 & -725 & -723 & -717 & -712 & -705 & -694 & -685 & -679 & -673 & -667 & -658 & -647 & -606 & -554 & -499 & -444\\
\multicolumn{2}{c}{hyd0073} & 8.2 & 8.3 & 8.7 & 9.0 & 9.4 & 10.3 & 11.4 & 12.5 & 13.7 & 14.9 & 16.1 & 17.5 & 20.9 & 24.5 & 28.1 & 31.8\\
6726 & 4.25 & -925 & -916 & -890 & -876 & -857 & -824 & -795 & -770 & -747 & -725 & -703 & -680 & -616 & -550 & -485 & -421\\
\multicolumn{2}{c}{hyd0077} & 6.3 & 6.6 & 7.3 & 7.7 & 8.3 & 9.5 & 10.9 & 12.5 & 14.2 & 15.9 & 17.7 & 19.5 & 24.1 & 29.0 & 33.8 & 38.6\\
4480 & 4.00 & -318 & -324 & -335 & -339 & -343 & -351 & -361 & -373 & -386 & -397 & -404 & -408 & -403 & -382 & -349 & -309\\
\hline\noalign{\smallskip}
\end{tabular}
\end{table*}

We used the library of synthetic spectra presented by \citet{Ludwig17} to
derive theoretical estimates of convective velocity shifts expected in the
spectra of the observed stars. For the Sun, \citet{2009MmSAI..80..622A} showed
that the accuracy of the predicted shifts is in the order of $\pm
70\,\mathrm{m\,s^{-1}}$ which we take as indicative for all model spectra
applied here. To stay close to the observational procedure, we determined the
convective line shifts by cross-correlating the synthetic spectra with the
weighted binary G2-mask of the \HARPS\ pipeline \citep{2002A&A...388..632P} covering a
wavelength range of 375--680\pun{nm}. No additional weighting was
considered. The approach is only an approximation of the actual procedure
applied in the \HARPS\ reduction pipeline since in the pipeline the
correlation is performed for each echelle order separately, and weighted by
the wavelength-dependent observed stellar flux. However, the
accuracy achieved by our simplified approach is likely sufficient for the present purpose.

The instrumental resolution and stellar rotation (in particular for F-type
dwarfs) lead to an additional broadening of a spectrum which is not part of
model syntheses as such. The additional broadening influences the position of
the peak of the cross-correlation function. We modelled this by convolving the
synthetic spectra with Gaussians $\propto \exp(-[\Delta\,v/\xi]^2)$ with $\xi$
ranging from 0 to 20\pun{km s$^{-1}$} (see Table~\ref{t:vconv}) which was
intended to mimic the total effect of instrumental and rotational
broadening. For a resulting cross-correlation function we calculated the
position of the maximum by a parabolic fit to the four uppermost points. The
width of the peak we obtained from a fit of a Gaussian to the peak and
the surrounding roughly constant base level. In particular at high resolution the
Gaussian was not always providing a good fit to the peak. We ignored this fact
since the \HARPS\ pipeline follows a similar procedure.

The convective shifts were calculated by interpolation in the data given in
Table~\ref{t:vconv}. This was done by a two step procedure. In a first step, we
interpolated the convective line shifts \vconv\ of all models to the FWHM of
the cross-correlation function measured for a particular object. The 1D
interpolations could be performed without problems since we had chosen a
rather dense grid of broadenings ($\xi$). The only caveat here was related to
stars with narrow cross-correlation peak ($\mbox{FWHM}\lesssim 8\,\mbox{km
 s$^{-1}$}$). In this case some model spectra have cross-correlation peaks
that are already broader than the targeted width without additional
broadening. This is due to their intrinsically high thermal and/or
micro-/macroturbulent broadening. In such cases we took the boundary value for
zero additional broadening as velocity shift. Over the range of broadenings
the fitting gave rms deviations always smaller than 30\pun{m s$^{-1}$} with a
maximum difference of (-)74\pun{m s$^{-1}$}. No model(s) could be identified
as outlier(s). In view of the precision expected for individual models we
considers this is satisfactory.

\begin{figure}
\begin{center}
\includegraphics[height=\columnwidth, angle=90]{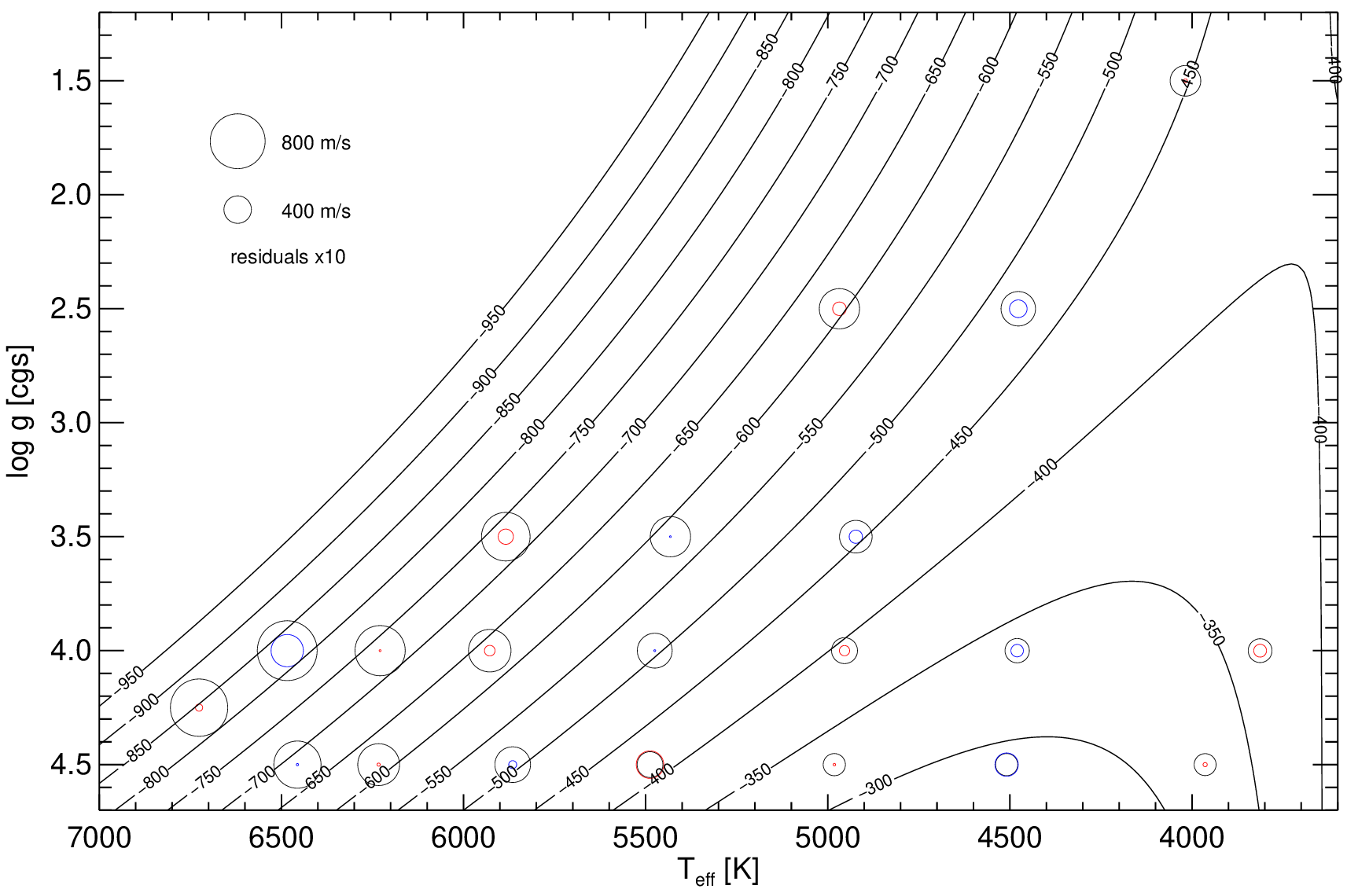}
\caption{Predicted convective blueshifts as estimated by cross-correlation for
 a FWHM of the cross-correlation peak of 10\,\pun{km s$^{-1}$}. Black
circles depict the underlying models in the \Teff-$\log g$-plane, their
diameters are proportional to the magnitude of the convective shifts. Contour
lines show a two-dimensional polynomial fit according to Eq.~(\ref{e:biqfit})
to the simulation results. Colored circles indicate residuals
$\vconv(\mathrm{fit})-\vconv(\mathrm{model})$, with red meaning positive and blue
negative deviations.\label{f:vconv}}
\end{center}
\end{figure}

In a second step, we fitted a biquadratic polynomial in $\log\Teff$ and
\logg\ to the 21 interpolated velocities. This was done in order to obtain a
higher accuracy by averaging over several models. The fitting polynomial had
the form
\begin{equation}
\vconv = a_1 + a_2\,t + a_3\,g + a_4\,t^2 + a_5\,t\,g,
\label{e:biqfit}
\end{equation}
where $t\equiv\log\Teff$ and $g\equiv\logg$. The five fitting coefficient
$a_i$ were obtained by solving the linear least-squares problem with equal
weights for all 21 data points. We left out the term proportional to $g^2$
since the term made the functional dependence of $\vconv(t,g)$ appear
unplausible at high levels of broadening. Our assumption -- or prejudice if
you wish -- was that to first order the functional dependence of
\vconv\ should follow the entropy jump \citep[see, e.g.,][for a discussion of
 the entropy jump]{1999A&A...346..111L} present in the model atmospheres
underlying the synthetic spectra. We finally evaluated \vconv\ of a star for
the measured \Teff\ and \logg\ from the fit. Figure~\ref{f:vconv} provides an
illustration of the outcome of the interpolation procedure for width of the
cross-correlation peak of 10\pun{km s$^{-1}$}.

\begin{figure}
\begin{center}
\includegraphics[height=\columnwidth, angle=90]{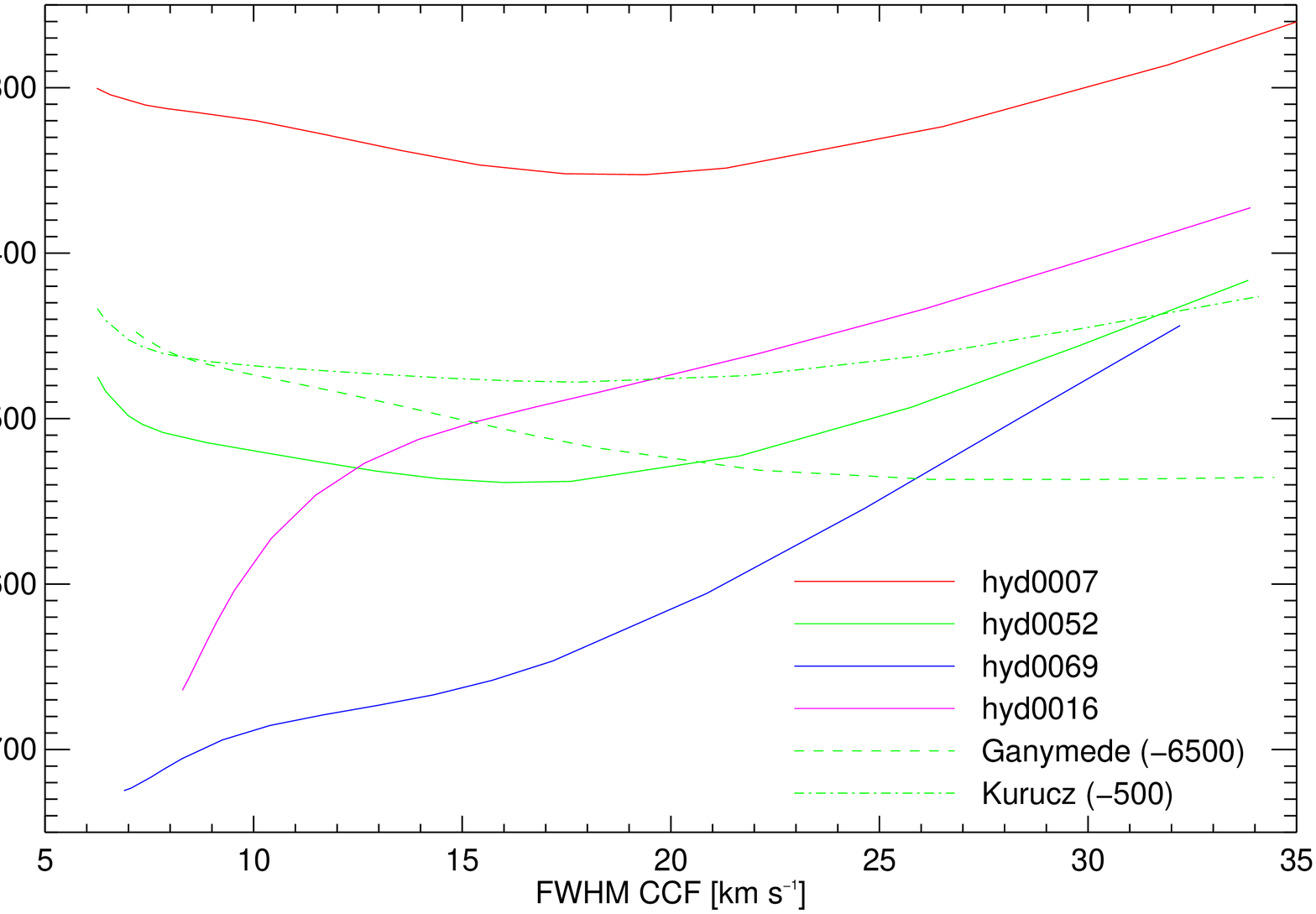}
\caption{Convective shifts as a function of the FWHM of the cross-correlation
 peak. Four model spectra are plotted representative of various
 stellar types (see also Table~\ref{t:vconv}): ``hyd0007'' K-dwarf, ``hyd0052''
 G-dwarf, ``hyd0069'' F-dwarf, ``hyd0016'' red giant. The curves labelled
 ``Kurucz'' and ``Ganymede'' are for two solar spectra discussed in the
 text; their curves were arbitrarily shifted by -500 and -6500\pun{m s$^{-1}$}, respectively.
 \label{f:resol}}
\end{center}
\end{figure}

Finally, we would like to comment on the dependence of the convective shifts
on spectral resolution. At first glance, one might think that the spectral
resolution has no influence since commonly considered broadening mechanisms
(instrumental broadening, micro-/macroturbulence, stellar rotation) broaden
spectral lines symmetrically. However, one has to keep in mind that spectral
lines are intrinsically asymmetric in stellar spectra, shifted differently
depending on strength, and often blended. A change of spectral resolution
alters their relative contributions contributions, and ultimately leads to a
shift of the maximum of the cross-correlation
function. \citet{1979AJ.....84.1511T} conduct an analytic analysis of the
cross-correlation technique and demonstrate a dependence of the measured
RV on the width of the cross-correlation peak. Quantitatively,
the effect depends on the particular circumstances. Figure~\ref{f:resol} gives
an illustration for the synthetic spectra and two solar spectra, namely those of
Ganymede and from the atlas of \citet{Kurucz84}. It is
apparent that in all cases there is a dependence, however it is weakest for
the empirical atlases. Since the \HARPS\ G2-mask was tailored after the solar
spectrum one might conjecture that mismatches between line positions and
strengths are smallest here, and that might lead to the weak
dependence. Conversely, one might take this as a warning that significant
mismatches between mask of observed spectrum might enhance the resolution
dependence. In any case, we tried to carefully include the resolution effect
in our estimates of the convective line shifts.


\section{Additional astrometric data}
\label{a:addastro}

Table \ref{table:addastro} provides some additional astrometric data for the cross-match between our final sample of 71 objects and the {\em Gaia} Archive (subsample of 65 objects).
The table contains the distance from the cluster centre, $|\mathbf{r} - \mathbf{r}_c|$, and tangential components in the $\alpha$ and $\delta$ directions for the astrometric velocity vectors computed from G17 data. The tangential components are given by $V_{\alpha} = k \mu_{\alpha*} / \pi$ and $V_{\delta} = k \mu_{\delta} / \pi$, where $\mu_{\alpha*} = \mu_{\alpha} \cos{\delta}$ and $\mu_{\delta}$ are the proper motions given in mas~yr$^{-1}$ and $k = 4.74047$~km~yr~s$^{-1}$ is the factor that transforms from AU~yr$^{-1}$ to km~s$^{-1}$ (e.g., P98).


\begin{table*}
 \caption{Additional astrometric data.}
 \label{table:addastro}
 {\centering 
 \begin{tabular}{ccccccccccc}
\toprule
\# & HIP & $|\mathbf{r} - \mathbf{r}_c|$ & $V_{\alpha}$ & $V_{\delta}$ & & \# & HIP & $|\mathbf{r} - \mathbf{r}_c|$ & $V_{\alpha}$ & $V_{\delta}$ \\
 & & (pc) & (km s$^{-1}$) & (km s$^{-1}$) & & & & (pc) & (km s$^{-1}$) & (km s$^{-1}$) \\
\midrule
2 & 16529 & 12.08 $\pm$ 0.36 & 32.317 $\pm$ 0.354 & -7.852 $\pm$ 0.086 & & 43 & 19441 & 12.79 $\pm$ 0.39 & 26.129 $\pm$ 0.236 & -0.920 $\pm$ 0.022 \\
4 & 18327 & 8.51 $\pm$ 0.43 & 28.400 $\pm$ 0.266 & -4.712 $\pm$ 0.046 & & 44 & 19808 & 4.13 $\pm$ 1.14 & 25.346 $\pm$ 0.716 & -4.523 $\pm$ 0.130 \\
6 & 19148 & 4.28 $\pm$ 0.36 & 26.418 $\pm$ 0.293 & -4.357 $\pm$ 0.050 & & 45 & 19834 & 3.39 $\pm$ 0.33 & 25.865 $\pm$ 0.296 & -4.141 $\pm$ 0.068 \\
8 & 19781 & 3.79 $\pm$ 0.40 & 24.700 $\pm$ 0.331 & -3.758 $\pm$ 0.052 & & 46 & 19862 & 2.98 $\pm$ 0.68 & 25.799 $\pm$ 0.355 & -4.966 $\pm$ 0.075 \\
9 & 19796 & 5.90 $\pm$ 0.44 & 25.552 $\pm$ 0.275 & -1.179 $\pm$ 0.016 & & 47 & 20357 & 3.29 $\pm$ 0.57 & 24.164 $\pm$ 0.419 & -4.436 $\pm$ 0.077 \\
10 & 20146 & 1.74 $\pm$ 0.20 & 25.318 $\pm$ 0.313 & -6.221 $\pm$ 0.078 & & 48 & 20527 & 4.45 $\pm$ 0.55 & 23.575 $\pm$ 0.252 & -3.862 $\pm$ 0.055 \\
12 & 20480 & 4.77 $\pm$ 0.56 & 23.418 $\pm$ 0.293 & -8.940 $\pm$ 0.113 & & 51 & 20762 & 3.51 $\pm$ 0.24 & 23.845 $\pm$ 0.276 & -3.765 $\pm$ 0.052 \\
13 & 20492 & 2.11 $\pm$ 0.40 & 24.018 $\pm$ 0.274 & -3.949 $\pm$ 0.051 & & 52 & 20827 & 3.19 $\pm$ 0.49 & 23.375 $\pm$ 0.310 & -4.478 $\pm$ 0.061 \\
14 & 20557 & 5.41 $\pm$ 0.49 & 23.819 $\pm$ 0.328 & -9.294 $\pm$ 0.128 & & 53 & 21138 & 2.02 $\pm$ 0.44 & 22.646 $\pm$ 0.267 & -5.224 $\pm$ 0.069 \\
16 & 20826 & 4.34 $\pm$ 0.30 & 24.105 $\pm$ 0.274 & -2.733 $\pm$ 0.032 & & 54 & 21256 & 4.73 $\pm$ 0.36 & 22.336 $\pm$ 0.265 & -9.538 $\pm$ 0.114 \\
17 & 20850 & 2.86 $\pm$ 0.23 & 23.240 $\pm$ 0.265 & -4.312 $\pm$ 0.052 & & 55 & 21261 & 2.20 $\pm$ 0.53 & 22.578 $\pm$ 0.271 & -7.616 $\pm$ 0.094 \\
19 & 20949 & 13.27 $\pm$ 1.46 & 22.297 $\pm$ 0.514 & -13.055 $\pm$ 0.301 & & 56 & 21723 & 5.73 $\pm$ 0.39 & 20.787 $\pm$ 0.212 & -3.275 $\pm$ 0.041 \\
20 & 20978 & 2.13 $\pm$ 0.59 & 22.631 $\pm$ 0.258 & -5.991 $\pm$ 0.070 & & 57 & 22177 & 11.46 $\pm$ 0.27 & 19.972 $\pm$ 0.230 & 2.846 $\pm$ 0.044 \\
21 & 21099 & 2.70 $\pm$ 0.19 & 23.244 $\pm$ 0.252 & -8.575 $\pm$ 0.094 & & 58 & 22253 & 10.21 $\pm$ 1.10 & 20.371 $\pm$ 0.333 & -10.645 $\pm$ 0.174 \\
22 & 21741 & 13.91 $\pm$ 1.04 & 21.206 $\pm$ 0.286 & -10.789 $\pm$ 0.146 & & 59 & 22271 & 8.46 $\pm$ 0.34 & 25.545 $\pm$ 0.357 & -18.881 $\pm$ 0.263 \\
23 & 22380 & 4.73 $\pm$ 0.22 & 18.475 $\pm$ 0.219 & -8.153 $\pm$ 0.098 & & 60 & 22654 & 6.42 $\pm$ 0.60 & 18.927 $\pm$ 0.216 & -8.410 $\pm$ 0.100 \\
24 & 22422 & 4.63 $\pm$ 0.34 & 19.695 $\pm$ 0.227 & -5.790 $\pm$ 0.068 & & 61 & 23312 & 13.71 $\pm$ 0.28 & 17.892 $\pm$ 0.223 & 2.455 $\pm$ 0.040 \\
25 & 22566 & 15.10 $\pm$ 1.31 & 18.970 $\pm$ 0.323 & -5.244 $\pm$ 0.090 & & 62 & 13806 & 19.09 $\pm$ 0.35 & 36.991 $\pm$ 0.409 & -8.925 $\pm$ 0.099 \\
26 & 23069 & 10.45 $\pm$ 0.80 & 18.037 $\pm$ 0.251 & -4.664 $\pm$ 0.068 & & 63 & 13976 & 26.74 $\pm$ 0.15 & 36.693 $\pm$ 0.228 & 2.367 $\pm$ 0.016 \\
27 & 23498 & 10.26 $\pm$ 0.75 & 16.694 $\pm$ 0.253 & -4.544 $\pm$ 0.070 & & 64 & 19786 & 4.71 $\pm$ 0.40 & 26.017 $\pm$ 0.340 & -2.959 $\pm$ 0.041 \\
28 & 23750 & 10.35 $\pm$ 0.86 & 16.268 $\pm$ 0.246 & -7.864 $\pm$ 0.119 & & 65 & 19793 & 6.00 $\pm$ 0.20 & 25.597 $\pm$ 0.290 & -10.319 $\pm$ 0.117 \\
29 & 24923 & 14.92 $\pm$ 0.73 & 13.646 $\pm$ 0.218 & -3.556 $\pm$ 0.059 & & 66 & 19934 & 6.41 $\pm$ 0.55 & 25.365 $\pm$ 0.320 & -9.306 $\pm$ 0.118 \\
31 & 15300 & 16.29 $\pm$ 0.33 & 33.491 $\pm$ 0.325 & -8.819 $\pm$ 0.092 & & 67 & 20082 & 2.54 $\pm$ 0.53 & 25.392 $\pm$ 0.258 & -5.003 $\pm$ 0.060 \\
32 & 15563 & 19.52 $\pm$ 0.25 & 33.905 $\pm$ 0.254 & 0.887 $\pm$ 0.017 & & 68 & 20130 & 2.93 $\pm$ 0.21 & 24.607 $\pm$ 0.317 & -7.964 $\pm$ 0.103 \\
33 & 15720 & 19.47 $\pm$ 0.27 & 32.938 $\pm$ 0.271 & -9.616 $\pm$ 0.082 & & 70 & 20485 & 5.64 $\pm$ 0.69 & 24.442 $\pm$ 0.319 & -5.176 $\pm$ 0.071 \\
35 & 17766 & 14.22 $\pm$ 0.38 & 30.414 $\pm$ 0.280 & 1.127 $\pm$ 0.021 & & 71 & 20563 & 1.26 $\pm$ 0.68 & 24.161 $\pm$ 0.455 & -6.674 $\pm$ 0.127 \\
36 & 18018 & 10.26 $\pm$ 0.42 & 29.452 $\pm$ 0.348 & -9.288 $\pm$ 0.111 & & 72 & 20577 & 1.66 $\pm$ 0.68 & 24.227 $\pm$ 0.283 & -5.365 $\pm$ 0.063 \\
37 & 18322 & 7.37 $\pm$ 0.45 & 28.505 $\pm$ 0.318 & -2.228 $\pm$ 0.030 & & 73 & 20741 & 1.21 $\pm$ 1.26 & 23.514 $\pm$ 0.505 & -6.067 $\pm$ 0.130 \\
38 & 18946 & 5.20 $\pm$ 0.19 & 27.761 $\pm$ 0.353 & -7.594 $\pm$ 0.100 & & 74 & 20815 & 1.33 $\pm$ 0.39 & 23.603 $\pm$ 0.436 & -5.340 $\pm$ 0.099 \\
39 & 19082 & 4.76 $\pm$ 0.25 & 27.399 $\pm$ 0.474 & -7.328 $\pm$ 0.131 & & 75 & 20899 & 0.63 $\pm$ 0.52 & 23.815 $\pm$ 0.306 & -6.225 $\pm$ 0.080 \\
40 & 19207 & 4.61 $\pm$ 0.57 & 26.401 $\pm$ 0.331 & -4.203 $\pm$ 0.057 & & 76 & 20951 & 1.47 $\pm$ 0.44 & 23.227 $\pm$ 0.255 & -7.097 $\pm$ 0.080 \\
41 & 19263 & 3.99 $\pm$ 0.43 & 26.433 $\pm$ 0.297 & -5.390 $\pm$ 0.065 & & 77 & 21317 & 2.02 $\pm$ 0.22 & 22.045 $\pm$ 0.245 & -5.830 $\pm$ 0.066 \\
42 & 19316 & 5.47 $\pm$ 0.52 & 25.932 $\pm$ 0.332 & -2.361 $\pm$ 0.042 & & & & & & \\
\bottomrule
 \end{tabular} \\}
\end{table*}

\bsp	
\label{lastpage}
\end{document}